\documentclass[useAMS,usenatbib]{mn2e}
\usepackage{graphicx}
\usepackage[fleqn]{amsmath}
\usepackage{threeparttable}
\usepackage{ulem}

\newcommand{\aap}{A\&A}
\newcommand{\wmhzsr}{$\mathrm{W}\,\mathrm{m^{-2}}\,\mathrm{Hz^{-1}}\,\mathrm{sr^{-1}}$}

\title[Calibration of the SPIRE FTS]{Calibration of the {\it Herschel} SPIRE Fourier Transform Spectrometer \thanks{{\it Herschel} is an ESA space observatory with science instruments provided by European-led Principal Investigator consortia and with important participation from NASA.}}
\author[B. M. Swinyard et al.]{\noindent B. M. Swinyard$^{1, 2}$\thanks{E-mail: bruce.swinyard@stfc.ac.uk},
E. T. Polehampton$^{2, 3}$,
R. Hopwood$^{4}$,
I. Valtchanov$^{6}$,
\newauthor
N. Lu$^{7}$,
T. Fulton$^{8, 3}$,
D. Benielli$^{9}$,
P. Imhof$^{8, 3}$,
N. Marchili$^{13}$,
J.-P. Baluteau$^{9}$,
\newauthor
G. J. Bendo$^{12}$,
M. Ferlet$^{2}$,
M. J. Griffin$^{11}$,
T. L. Lim$^{2}$,
G. Makiwa$^{3}$, 
D. A. Naylor$^{3}$,
\newauthor
G. S. Orton$^{14}$,
A. Papageorgiou$^{11}$,
C. P. Pearson$^{2, 5}$,
B. Schulz$^{7}$,
S. D. Sidher$^{2}$,
\newauthor
L. D. Spencer$^{11}$,
M. H. D. van der Wiel$^{3}$,
R. Wu$^{10}$\\
$^{1}$Department of Physics and Astronomy, University College London, Gower St, London WC1E 6BT, UK \\
$^{2}$RAL Space, Rutherford Appleton Laboratory, Chilton, Didcot, Oxfordshire, OX11 0QX, UK \\
$^{3}$Institute for Space Imaging Science, University of Lethbridge, 4401 University Drive, Lethbridge, Alberta, T1J 1B1, Canada \\
$^{4}$Department of Physics, Imperial College London, Prince Consort Road, London SW7 2AZ, UK \\
$^{5}$Department of Physical Sciences, The Open University, Milton Keynes, MK7 6AA, UK\\
$^{6}$Herschel Science Centre, European Space Astronomy Centre, ESA, 28691 Villanueva de la Ca\~nada, Spain\\
$^{7}$NASA Herschel Science Center, MS 100-22, California Institute of Technology, Pasadena, CA 91125, USA\\
$^{8}$Blue Sky Spectroscopy, Lethbridge, AB, T1J 0N9, Canada\\
$^{9}$Laboratoire d'Astrophysique de Marseille - LAM, Universit\'e d'Aix-Marseille \& CNRS, UMR7326, 38 rue F. Joliot-Curie, 13388 Marseille\\
Cedex 13, France \\
$^{10}$Laboratoire AIM, CEA/DSM - CNRS - Irfu/Service d'Astrophysique, CEA Saclay, 91191 Gif-sur-Yvette, France\\
$^{11}$School of Physics and Astronomy, Cardiff University, The Parade, Cardiff, CF24  3AA, UK\\
$^{12}$UK ALMA Regional Centre Node, Jodrell Bank Centre for Astrophysics, School of Physics and Astronomy, University of Manchester, \\
Oxford Road, Manchester M13 9PL, UK\\
$^{13}$Dipartimento di Fisica e Astronomia, Universit\`a di Padova, I-35131 Padova, Italy\\
$^{14}$Jet Propulsion Laboratory, California Institute of Technology, 4800 Oak Grove Drive, Pasadena, CA 91109, USA}

\begin{document}

\date{Accepted.. Received..; in original form ..}

\pagerange{\pageref{firstpage}--\pageref{lastpage}} \pubyear{2013}

\maketitle

\label{firstpage}

\begin{abstract}
The {\it Herschel} SPIRE instrument consists of an imaging photometric camera and an imaging Fourier Transform Spectrometer (FTS), both operating over a frequency range of $\sim$450--1550\,GHz. In this paper, we briefly review the FTS design, operation, and data reduction, and describe in detail the approach taken to relative calibration (removal of instrument signatures) and absolute calibration against standard astronomical sources. The calibration scheme assumes a spatially extended source and uses the {\it Herschel} telescope as primary calibrator. Conversion from extended to point-source calibration is carried out using observations of the planet Uranus.  The model of the telescope emission is shown to be accurate to within 6\% and repeatable to better than 0.06\% and, by comparison with models of Mars and Neptune, the Uranus model is shown to be accurate to within 3\%. Multiple observations of a number of point-like sources show that the repeatability of the calibration is better than 1\%, if the effects of the satellite absolute pointing error (APE) are corrected.  The satellite APE leads to a decrement in the derived flux, which can be up to $\sim$10\% (1\,$\sigma$) at the high-frequency end of the SPIRE range in the first part of the mission, and $\sim$4\% after {\it Herschel} operational day 1011.  The lower frequency range of the SPIRE band is unaffected by this pointing error due to the larger beam size.  Overall, for well-pointed, point-like sources, the absolute flux calibration is better than 6\%, and for extended sources where mapping is required it is better than 7\%.
\end{abstract}



\section{Introduction}\label{sect_intro}

The Spectral and Photometric REceiver \citep[SPIRE][]{griffin2010} is one of three focal plane instruments which operated on board the ESA {\it Herschel} Space Observatory \citep[{\it Herschel};][]{pilbratt10} between May 2009 and April 2013. It contains an imaging photometric camera and an imaging Fourier Transform Spectrometer (FTS), with both sub-instruments using arrays of bolometric detectors operating at $\sim$300~mK \citep{turner2001} and feedhorn focal-plane optics giving sparse spatial sampling over an extended field of view \citep{dohlen2000}. This paper details the calibration scheme adopted for the SPIRE FTS, updating the early description by \citet{swinyard2010}.

\begin{figure}
\centering
\includegraphics[width=\hsize]{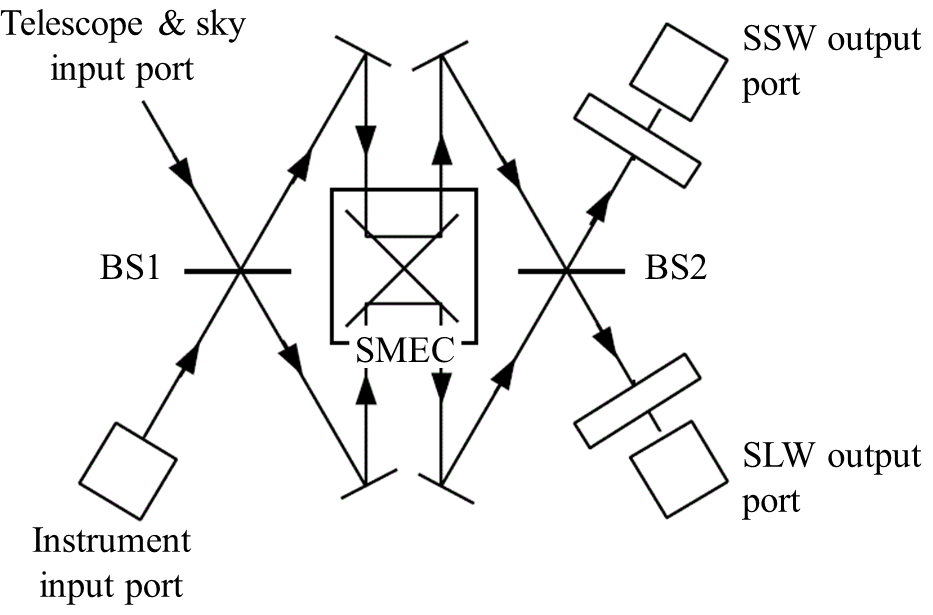}
\caption{The optical layout of the SPIRE FTS.}
\label{fts_design}
\end{figure}

The FTS uses two bolometer arrays of 19 and 37 detectors to provide spectral imaging over a nominal $\sim$2 arcminute field of view. The design of the SPIRE FTS \citep[][]{ade1999, dohlen2000, swinyard2010} is shown in Fig.~\ref{fts_design}: the incoming radiation from the telescope is divided into two beams by a beamsplitter (BS1). These beams are retro-reflected from back-to-back roof top mirrors (RT) mounted on a linear translation stage (the Spectrometer MEChanism; SMEC). The SMEC imparts an optical path difference (OPD) between the two beams dependent on the mirror position, and a second beam splitter (BS2) recombines the light to form an interference pattern that is focused onto the detector arrays. There are no significant spectral features within the band of the beam splitters \citep{ade1999}.

\begin{figure}
\centering
\includegraphics[width=\hsize]{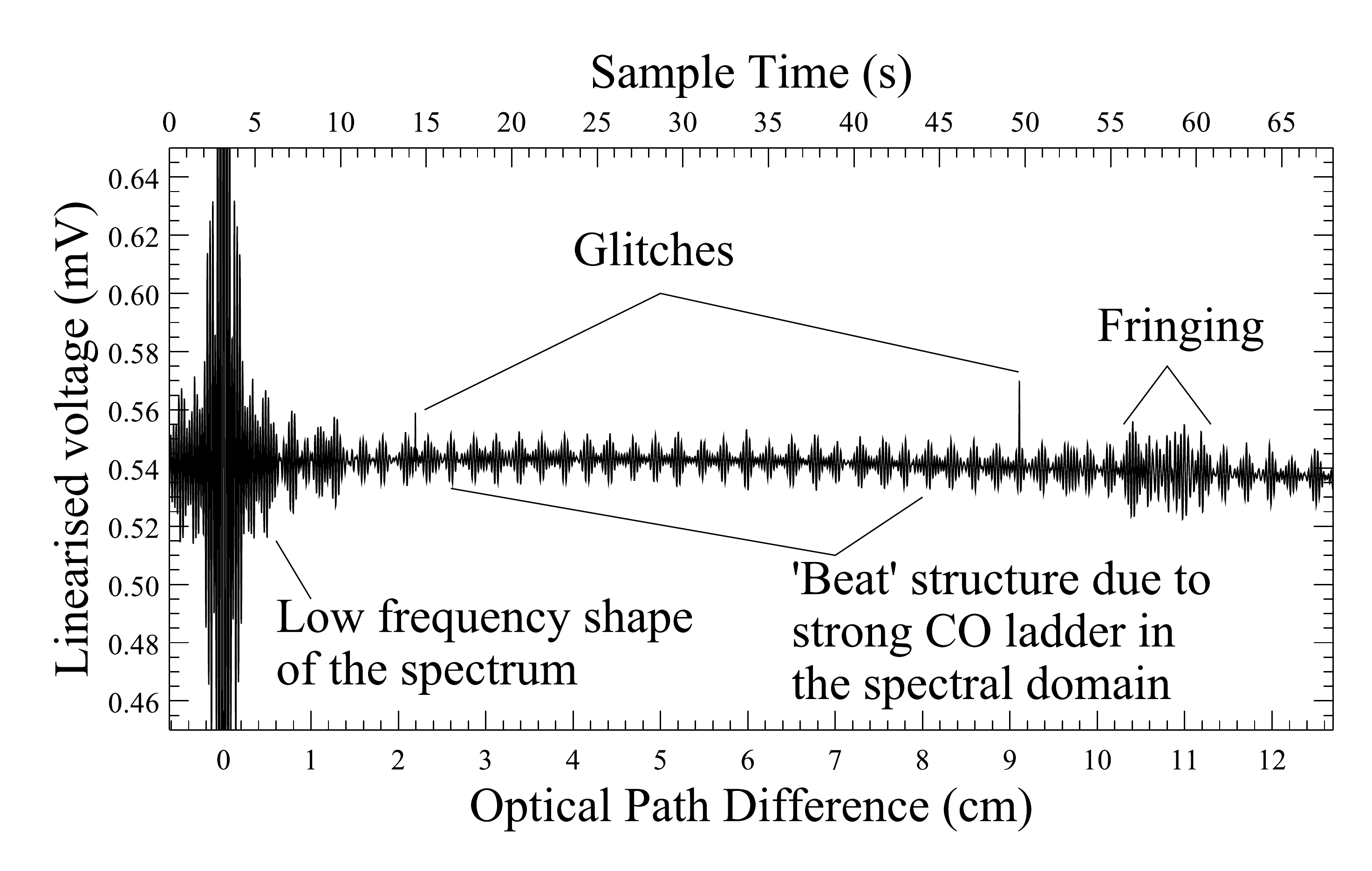}
\includegraphics[width=\hsize]{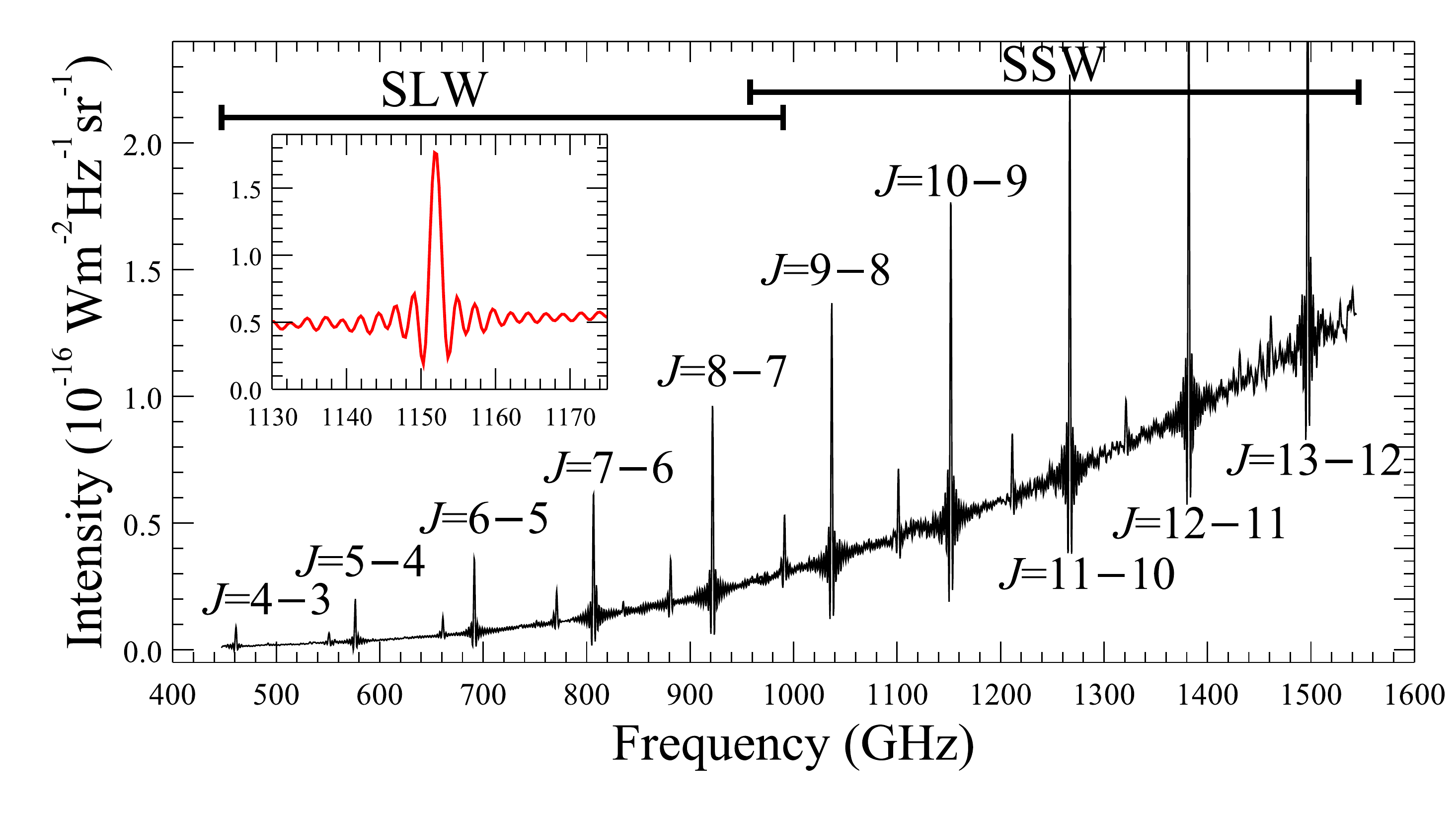}
\caption{Top: Typical measured interferogram from the SPIRE FTS for an astronomical source with strong $^{12}$CO lines. The main features of the interferogram are highlighted. Bottom: Final spectrum for this interferogram showing the SSW and SLW bands and with transitions of $^{12}$CO labelled. The insert shows a zoom around the $^{12}$CO $J$=10--9 line.}
\label{interferogram}
\end{figure}

The response of the detector system to the source intensity is measured as a function of the SMEC position and is hereinafter referred to as the ``interferogram''. The interferogram is the Fourier transform of the incident spectrum, as modified by the instrumental response and other instrumental effects. The use of two beam splitters in a Mach-Zehnder configuration \citep{mach,zehnder}, plus the back-to-back roof top arrangement, means that the imparted optical path difference is four times larger than the physical movement of the mechanism, making for a compact optical arrangement. The Mach-Zehnder configuration also provides natural spatial separation of the two input and two output ports always present in an FTS. In SPIRE the second input port is terminated on a cold radiative source (SCAL) which can be heated to provide a known radiation load onto the detectors \citep{hargrave2006}. The two output ports are chromatically separated to allow the instrument to cover a broad frequency range whilst maintaining close to optimal optical coupling to the detectors. There are two optimised arrays, called SSW (Spectrometer Short Wavelength, covering 959.3--1544~GHz) and SLW (Spectrometer Long Wavelength, covering 446.7--989.4~GHz)\footnote{The band limits may be expanded slightly in future versions of the pipeline.}. Fig.~\ref{interferogram} shows a typical interferogram for a source with strong spectral lines, and the equivalent spectrum observed using the high resolution mode of the instrument. The spectral resolution, defined as the distance from the peak to the first zero crossing of the instrumental line shape, is constant in frequency at $\sim$1.184\,GHz, equivalent to 230--800\,kms$^{-1}$ \citep{observersmanual}. Note that in general (except for very bright sources), the fringing shown in the top plot of Fig.~\ref{interferogram} is successfully removed by the calibration scheme as it is stable in both science and calibration observations.

The SCAL source was designed to increase the dynamic range of the detectors by nulling the modulated signal component of the interferogram. However, the total emission from the telescope and stray light were actually lower than the values used in the initial design of the SPIRE instrument, and the SCAL source was not needed and was therefore not actively heated.

\begin{figure}
\centering
\includegraphics[width=\hsize]{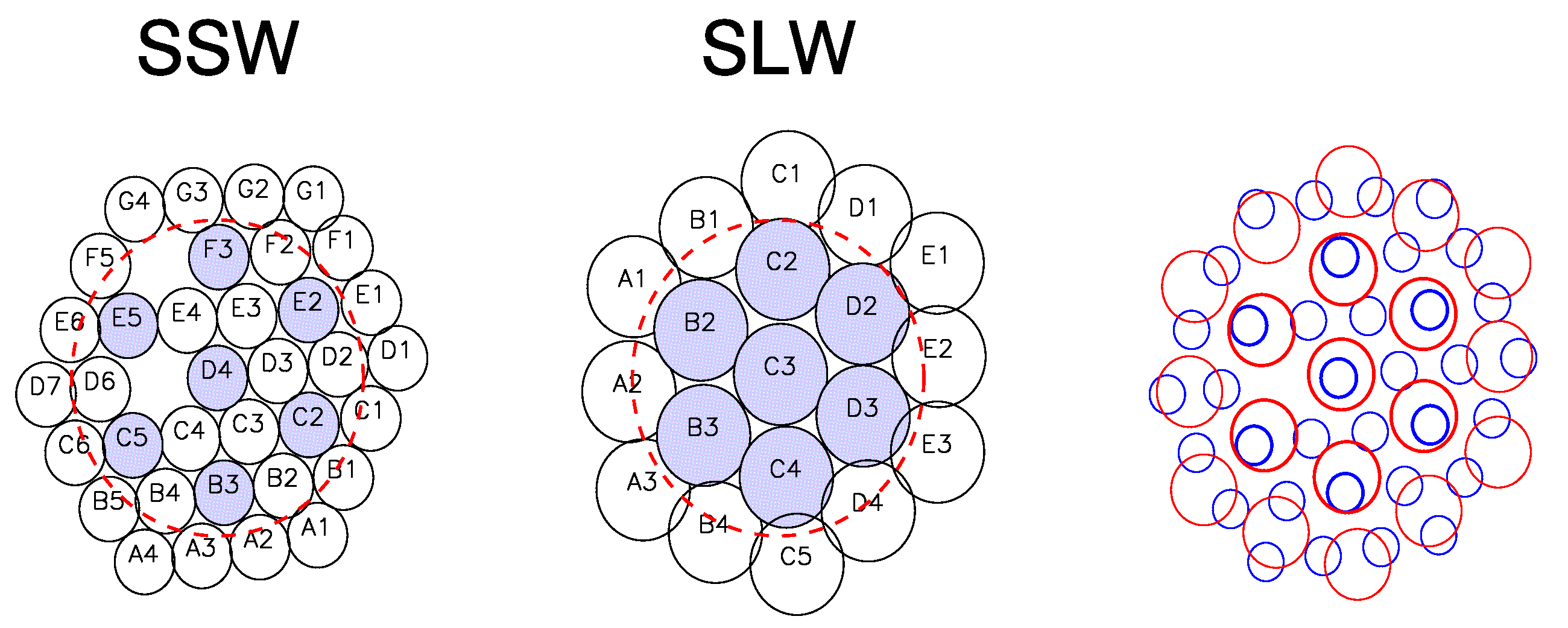}
\caption{A schematic view of the SPIRE FTS detector arrays showing the measured position of each detector. The right hand plot shows the two arrays as they appear on sky, with 19$^{\prime\prime}$ circles for SSW and 35$^{\prime\prime}$ for SLW. See main text for more details.}
\label{fts_array_footprint}
\end{figure}

A schematic view of the SPIRE FTS detector arrays is shown in Fig.~\ref{fts_array_footprint}, showing the relative positions of each detector as measured at the beginning of the mission. The detectors are arranged in a hexagonally close-packed pattern with the spacing between beam centres set to $\sim$33$^{\prime\prime}$ for SSW and $\sim$51$^{\prime\prime}$ for SLW, roughly equal to two beam widths. Vignetting and distortion within the optical design of the instrument increases away from the centre of each array, effectively limiting the nominal (unvignetted) field of view to $\sim$2$^{\prime}$. The nominal field of view is shown in Fig.~\ref{fts_array_footprint} as a dashed red line. The circles shaded in blue represent SSW and SLW detectors centred on the same sky positions and the gaps in the SSW array show the location of two dead detectors (SSWD5, SSWF4). The plot on the right in Fig.~\ref{fts_array_footprint} indicates the approximate full width at half maximum (FWHM) of the beam for each detector, and the overlap of the arrays on the sky, with 19$^{\prime\prime}$ circles for SSW and 35$^{\prime\prime}$ circles for SLW.

In this paper we describe the photometric, spectroscopic and spatial calibration of the SPIRE FTS in its nominal mode \citep[additional calibration needed for its bright-source mode is described by][]{lu}. In Section~\ref{sect_flux_cal} we describe the photometric calibration starting from the engineering data output from the detector electronics through to the derivation of calibrated spectra of astronomical objects; in Section~\ref{sect_fcfactors}, we describe the derivation of the flux conversion factors; in Section~\ref{sect_freq} we deal with the spectroscopic calibration and in Section~\ref{sect_beam} with the spatial response calibration. In Section~\ref{sect_accuracy} we summarise the accuracy and repeatability of the calibration, discuss caveats on the SPIRE data and consider aspects in which we expect to see improvements as our knowledge of the data improves. The calibration described has been implemented in the {\it Herschel} Interactive Processing Environment \citep[HIPE;][]{ott} Version 11 and a companion paper, Fulton et al. (in preparation), will detail how the procedures described here are put into practice in the SPIRE FTS data pipeline. Note that all errors in this paper are quoted as 1-sigma (1$\sigma$) limits.


\section{Flux calibration}\label{sect_flux_cal}

There are two steps in the SPIRE FTS processing pipeline that determine the absolute flux calibration: linearisation of the bolometer signal timeline, and absolute scaling of the spectrum into astronomical units using standard sources. These steps are described in the following sections.

\subsection{Linearisation of bolometer signals}

\begin{figure*}
\centering
\includegraphics[width=\hsize]{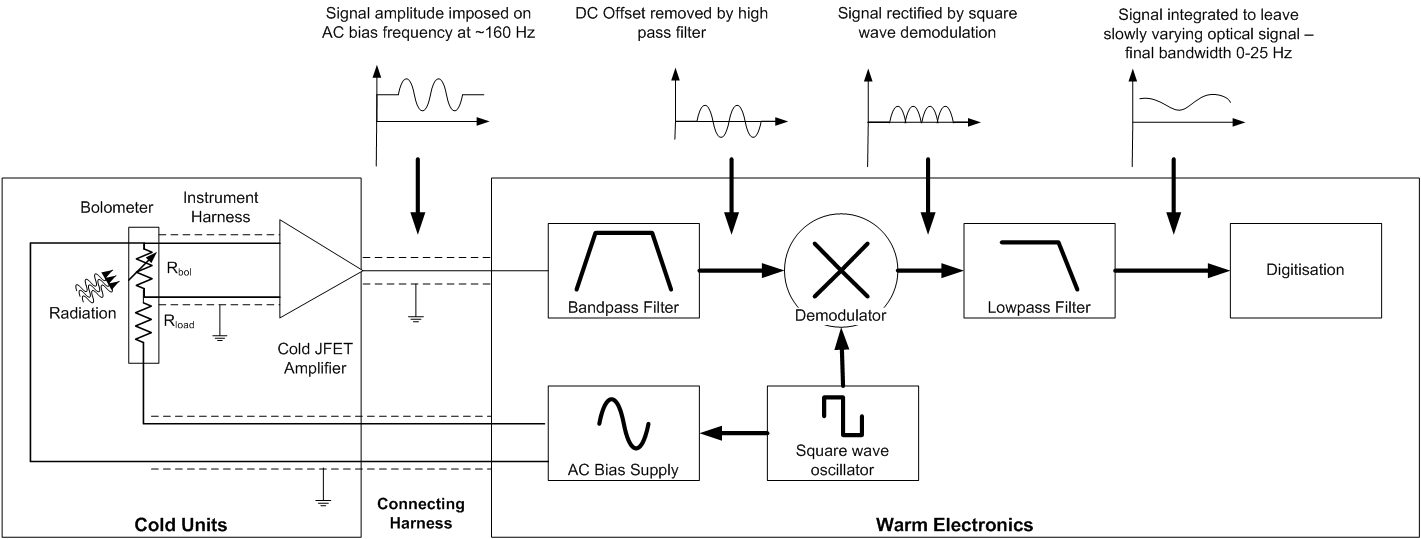}
\caption{Diagram showing the elements of the bolometer signal processing chain.}
\label{signal_chain}
\end{figure*}

The readout electronics for the SPIRE FTS bolometers are summarised in Fig.~\ref{signal_chain}. They consist of the cold junction field-effect transistors (JFETs), located in the cryostat close to the detectors, and the warm Detector Control Unit (DCU), located on the spacecraft Service Module. The cold and warm units are connected by the $\sim$5\,m long {\it Herschel} cryoharness. The DCU contains a separate lock-in amplifier (LIA) for each bolometer, and a 16-bit Analogue to Digital Converter (ADC). The dynamic range of the ADC output is set inside the electronics by automatically subtracting a constant DC offset level that is determined depending on the input power level \citep{griffin2010}. The offset level for each detector was measured and reset once, at the beginning of each observation, with the exception of mapping observations in bright-source mode, where they were measured and reset more frequently within the observation to account for source contrast \citep{observersmanual}.

The fundamental drive frequency for the bolometer AC bias is derived from a square-wave generator. The bias is provided from the warm electronics unit to the bolometer at the cold focal plane unit via a $\sim$5~m harness that is run through the {\it Herschel} cryostat. The signal from the bolometer is amplified at the cold focal plane unit using a source-follower JFET amplifier to reduce the output impedance, and therefore avoid phase and amplitude roll off down the long connecting harness.  However, the capacitance of the instrument harness connecting the bolometer to the JFET, in conjunction with the resistance to ground formed by the bolometer and its load-resistor, forms an RC circuit that does affect both the amplitude and phase of the signal provided to the lock-in amplifier in the warm electronics.  There is an additional fixed phase offset (i.e. not dependent on the bolometer resistance) due to the impedance of the connecting harness.  

The AC signals from the bolometer are synchronously demodulated using the square-wave generator. The phase between the square wave and the input signal is adjusted to provide the maximum signal after demodulation. This operation was performed at the start of the mission. However, as the power on the bolometer increases, its resistance falls and the signal and demodulation square wave move out of phase with each other by a small amount, reducing the effective gain in the circuit. A correction for this effect is implemented using the following method.

The root mean square (RMS) bolometer voltage is calculated from the ADC output and the offset level in two stages. First, the total gain of the DCU chain is applied. Next, the bolometer voltage, resistance and current are calculated by applying the gain due to the JFET and the cables. The cable gain accounts for the small change in phase described above, and so depends on the total resistance, which in turn depends on the final calculated bolometer resistance, and so an iteration is necessary. The iteration is performed until the fractional change in bolometer voltage is less than $10^{-4}$.

The DCU and JFET gains are contained in a calibration file derived from ground-based measurements. The cable gain, $G_{\rm cab}$, is calculated from
\begin{equation}
G_{\rm cab}=\sqrt{\left(\frac{1}{1+\omega_{cr}^{2}}\right)},
\end{equation}
where
\begin{equation}
\omega_{cr} = 2\pi\nu_{\rm bias}R_{\rm tot}C_{\rm H},
\end{equation}
$\nu_{\rm bias}$ is the frequency of the bolometer bias voltage, $R_{\rm tot}$ is the total resistance and $C_{\rm H}$ is the capacitance of the cables, which is $\sim$20~pF for the FTS. 

The iterative gain correction relies on a good adjustment of the detector bias to be ``in-phase'' with the square-wave generator at the beginning of the mission. This overall phase was monitored at intervals throughout the mission and found not to have varied significantly for the nominal mode of the FTS - i.e. the iteration described above was sufficient without any recourse to regular resetting of the overall phase. However, when very bright sources were observed using the bright-source mode, the phase setting and changes in phase between sources were important, and so an additional phase gain factor must be calculated explicitly. The bright-source mode calibration is described in detail by \citet{lu}. 

The RMS bolometer voltage only responds linearly to incoming radiation within a limited range of power, and therefore a correction for non-linearity is required. In a procedure similar to that adopted for the SPIRE photometer \citep[see][]{bendo}, the linearisation is carried out by integrating over the inverse bolometer (non-linear) response function, $f(V)$, between a fixed reference voltage, $V_{0}$, and the measured voltage, $V_{m}$,
\begin{equation}
S = \displaystyle\int\limits_{V_{0}}^{V_{m}}\!f(V)\,{\rm d}V\hspace{0.5cm}(\mathrm{V}),
\end{equation}
where $S$ is a measure of the optical load on the detector. For the spectrometer, this equation is normalised to the value of $f(V)$ at the reference voltage to give a signal that is proportional to the optical load on the detector. Although this quantity is a dimensionless proportional quantity, we refer to it as the linearised voltage, $V^{\prime}$,
\begin{equation}
V^{\prime} = \frac{1}{f(V_{0})}\displaystyle\int\limits_{V_{0}}^{V_{m}}\!f(V)\,{\rm d}V.
\end{equation}

In the same way as for the photometer, the normalised value of $f(V)$ can be approximated using
\begin{equation}
\frac{f(V)}{f(V_{0})} = K_1+\frac{K_2}{V-K_3 },
\end{equation}
where $K_1$, $K_2$ and $K_3$ are constants specific to each bolometer. For the nominal mode of the FTS, a bolometer model \citep{mather, sudiwala}, which is based on the bolometer thermometry measured in laboratory, and heat conductance parameters measured in flight \citep{hien2004}, is used to calculate the three $K$ parameters, and the linearised voltage is given by
\begin{equation}\label{eqn_linearisation}
V^{\prime}=K_1 (V_m-V_0 )+K_2\, \mathrm{ln}\left(\frac{V_m-K_3}{V_0-K_3 }\right).
\end{equation}

\begin{figure}
\centering
\includegraphics[trim = 10mm 130mm 15mm 60mm, clip, width=\hsize]{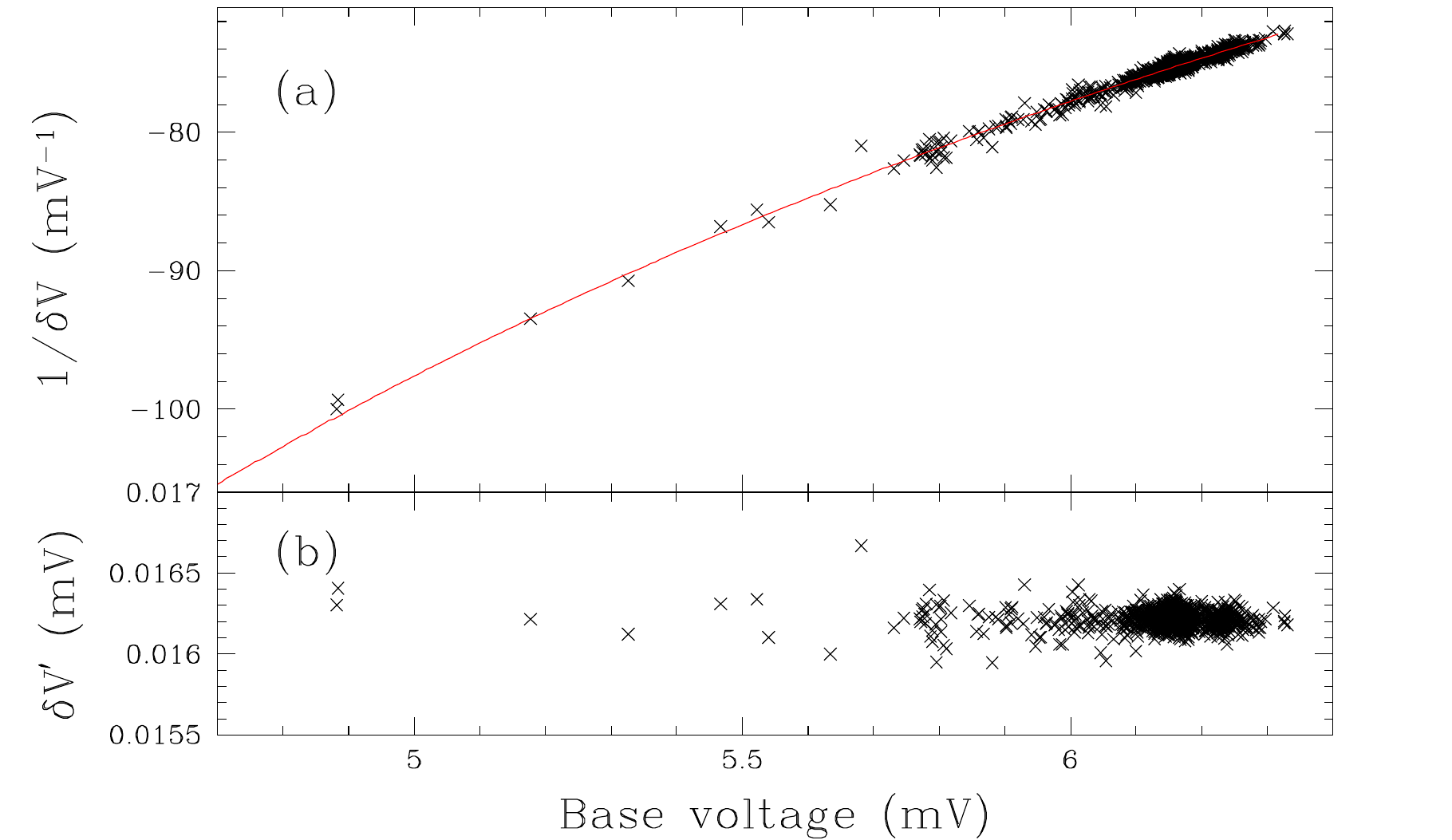}
\caption{The median voltage separation of the on and off cycles in individual PCAL flashes taken between ODs 209 and 1263 using detector SSWD4 as a function the observed median off voltage, where (a) uses the observed voltage separation ($\delta V$) while (b) the linearised counterpart ($\delta V$'). The red curve in (a) shows the adopted bolometer model-based nonlinearity curve normalised using the reference voltage, $V_0$. Note that the base voltage on the x-axis is as measured - i.e. before linearisation.}
\label{fig:pcal_flashes}
\end{figure}

The accuracy of the model-based approach has been checked by examining SPIRE photometer calibrator \citep[PCAL;][]{pisano} flashes, where a repeatable small change in detector power is provided by cycling the PCAL source on and off (on top of the astronomical and telescope background level). These flashes are carried out as part of every SPIRE observation. Fig.~\ref{fig:pcal_flashes} shows a compilation of all PCAL flashes between operational days\footnote{{\it Herschel} operational days are defined from the start of the mission on 14 May 2009} (ODs) 209 and 1263 for detector SSWD4. The linearised level separation of on and off cycles ($\delta V^{\prime}$) is consistent over a wide range of background levels with an RMS scatter of less than 1\%, with the other detectors showing a similar scatter.

The majority of observations were made with a stable base temperature for the detectors and the characterisation of the bolometer response described above works well. However, some observations made near the beginning of a SPIRE cooler cycle suffer from rapidly changing detector temperatures. These unstable conditions only have a significant effect on spectra of very faint sources (these observations still fall within the PCAL scatter quoted above) and, for certain detectors, can lead to a systematic loss of flux due to an over subtraction of the telescope emission. The pipeline will be expanded to correct this effect in a future version of HIPE.

The reference voltage does not affect the final spectral calibration, which depends only on the relative modulation around the interferogram baseline at spatial frequencies inside the optical band. This also means that any 1/$f$ noise, or large scale instrument drifts on a timescale much longer than one SMEC scan do not affect the final calibration. The interferogram baseline level is subtracted using a Fourier filter (Fulton et al., in preparation), removing any effect of the reference voltage $V_0$ which can be set to an arbitrary value. In practice, $V_0$ is set to the typical voltage measured on a dark area of sky.

Further pipeline steps are applied to the linearised bolometer signal timelines to correct for glitches (e.g. due to cosmic ray impacts) and clipped samples that hit the edge of the dynamic range of the ADC (Fulton et al., in preparation), before combining the signal and SMEC timelines to create the measured interferogram with signal as a function of mirror position. Any asymmetry about zero path difference (ZPD) is corrected using a phase correction algorithm which will be described in Fulton et al. (in preparation).

\subsection{Absolute scaling of the spectrum}

The linearised, baseline-subtracted interferogram represents a combination of optical power from the astronomical source, emission from the telescope (at 87--90~K\footnote{87~K corresponds to a black body with a peak at 5116~GHz}) and emission from inside the instrument (at $\sim$4.5~K\footnote{4.5~K corresponds to a black body with a peak at 265~GHz}). The interferogram signal at ZPD measures half of the total power across all frequencies in the band, the other half having been subtracted with the base level of the interferogram. Absolute flux calibration to convert to astronomical units is carried out on the spectrum obtained after applying a Fourier transform to this interferogram. The units of the uncalibrated spectrum are V\,GHz$^{-1}$, which we refer to as voltage density.

For all non-mapping observations, the standard pipeline scheme applies absolute flux calibration in two stages, to produce Level-1 and Level-2 products in the {\it Herschel} data structure. Firstly, voltage density is converted to intensity (surface brightness), $I_{\mathrm{ext}}$, which is appropriate for sources uniformly extended in the beam, and provides the Level-1 product, in units of \wmhzsr. Secondly, a point-source calibration is applied to convert $I_{\mathrm{ext}}$ to flux density, $F_{\mathrm{point}}$, providing the Level-2 product, in units of Jy. The observer is then able to select which of Level-1 or Level-2, if either, is most appropriate given the characteristics of the source. Corrections for partially extended sources that do not fit either assumption must be applied separately to the standard pipeline, and are described in \citet{wu}.

The spectrum, in units of linearised voltage density, can be expressed as
\begin{align}\label{eqn_ft}
V_{\rm obs} = {\rm FT}(V^{\prime}(t)) = R^*(\nu)I_{\rm ext}(\nu) + R_{\rm tel}(\nu) M_{\rm tel}(T, \nu)\nonumber
  \mspace{100mu}
  \notag\\
\hspace*{0.5cm}+ R_{\rm inst}(\nu) M_{\rm inst}(T, \nu)\hspace{0.5cm}(\mathrm{V\,GHz^{-1}}),
\end{align}
where $M_{\rm tel}$ and $M_{\rm inst}$ are the modelled intensities of the telescope and instrument respectively. The spectral calibration factors required to convert between voltage density and intensity are referred to as relative spectral response functions (RSRFs), although they also contain the absolute conversion between V\,GHz$^{-1}$ and \wmhzsr. The RSRFs for the source, telescope and instrument are $R^{*}$, $R_{\rm tel}$, $R_{\rm inst}$ respectively.

The telescope emission completely fills the SPIRE beam and is assumed to be uniform across the beam. Therefore, in order to calculate the astronomical source intensity, assuming that it is also fully and uniformly extended across the beam, $R^{*}$ is set equal to $R_{\rm tel}$,
\begin{equation}
\label{eqn_extended}
I_{\rm ext} = \frac{(V_{\rm obs} - M_{\rm inst}R_{\rm inst})}{R_{\rm tel}} - M_{\rm tel}\hspace{0.5cm}(\mathrm{W}\,\mathrm{m^{-2}}\,\mathrm{Hz^{-1}}\,\mathrm{sr^{-1}}),
\end{equation}
where $V_{\rm obs}$ is the observed voltage density spectrum defined in equation~\ref{eqn_ft}. The instrument and telescope RSRFs are different because the instrument contribution is dominated by the SCAL input port, whereas the telescope contribution enters only through the sky port. The models of the instrument and telescope emission, $M_{\rm inst}$ and $M_{\rm tel}$, and the derivation of the RSRFs are described in detail in Sections~\ref{sect.inst} and~\ref{sect.tele}. The intensity spectrum, $I_{\rm ext}$, forms the Level-1 data in the pipeline. For point sources, these Level-1 spectra are not particularly useful, but act as an intermediate step to the Level-2 data.

In the second step of the pipeline, the Level-1 data, $I_{\rm ext}$, are converted into flux density in units of Jy with a point-source calibration based on observations of Uranus,
\begin{equation}
\label{eqn_point}
F_{\rm point} = I_{\rm ext}\ C_{\rm point}\hspace{0.5cm}(\mathrm{Jy}),
\end{equation}
where $C_{\rm point}$ is a frequency dependent point-source conversion factor, described in more detail in Section~\ref{sect.uranus}. The resulting flux-density spectrum, $F_{\rm point}$, forms the Level-2 data in the pipeline.

The calibration is calculated independently for two epochs during the mission to account for a change in position of the internal Beam Steering Mirror (BSM) by 1.7$^{\prime\prime}$, carried out on 18th February 2012 ({\it Herschel} operational day 1011). This change in position was applied to move the BSM closer to the optical axis and therefore reduce uncertainties associated with {\it Herschel} pointing (see Section~\ref{sect:pointUn}). The results from the two epochs are consistent, and are therefore analysed together in the remainder of the paper.

\section{Derivation of the flux conversion factors}
\label{sect_fcfactors}

\subsection{Instrument model and RSRF\label{sect.inst}}

\begin{figure}
\centering
\includegraphics[width=\hsize]{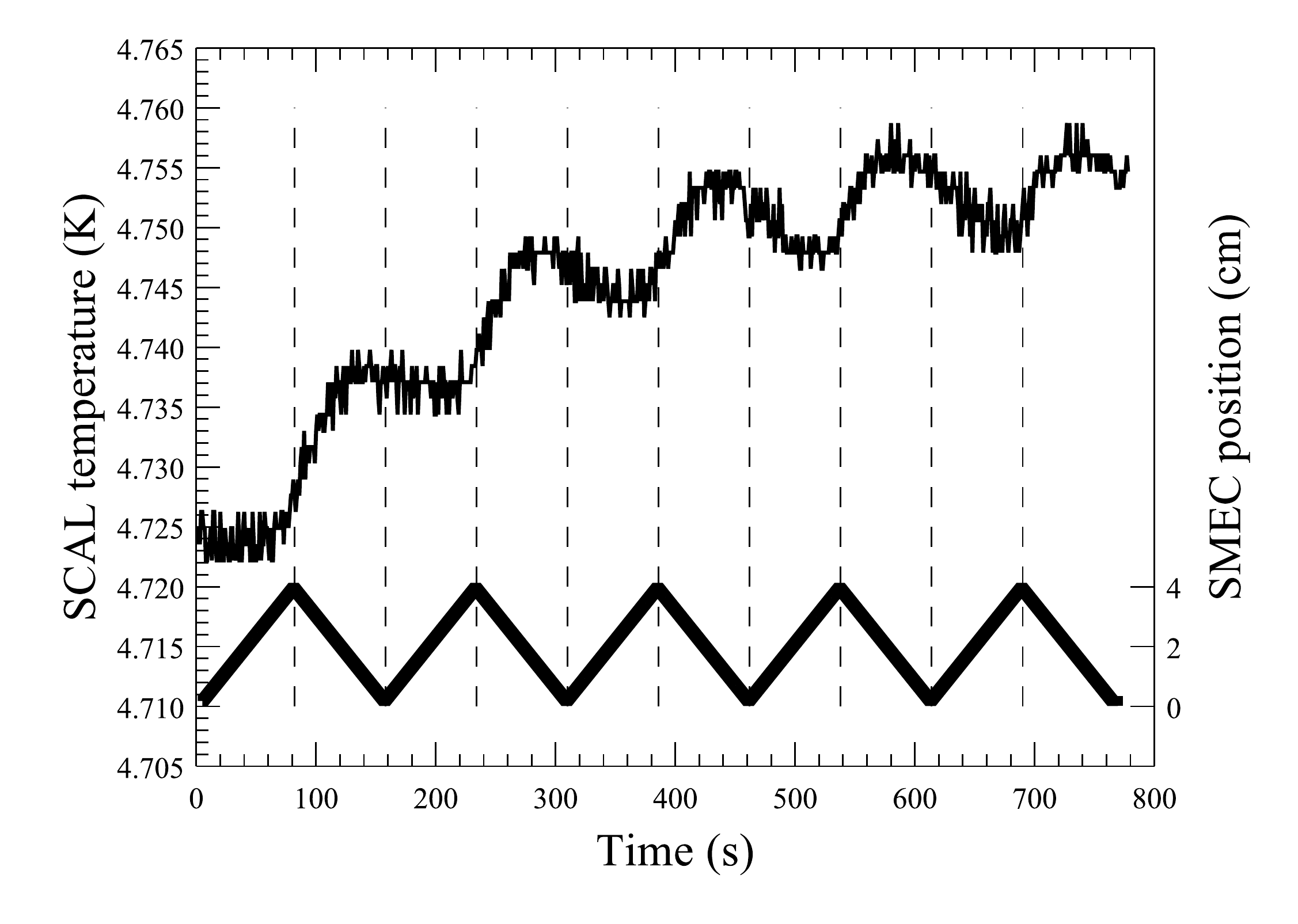}
\caption{The temperature of the SCAL source (the upper line in the plot) and the SMEC mechanical position (the triangular wave pattern at the bottom of the plot) during a typical high resolution observation. The start and end of each scan is indicated by the vertical dashed lines.}
\label{fig:scal_temp}
\end{figure}

The temperature of the instrument enclosure is maintained at approximately 4.5~K by the boil-off gas from the {\it Herschel} cryostat. However, during each observation, the SMEC scans backwards and forwards leading to local heating which causes additional power to enter the beam via the second port of the FTS. This port terminates on the SCAL calibration source which was not thermally controlled during the mission, and therefore propagates temperature changes directly into the beam. Fig.~\ref{fig:scal_temp} shows the measured temperature of the SCAL source compared to the SMEC position for a typical observation, making it clear that the instrument emission must be modelled separately for each SMEC scan. As described in \citet{fultonExpAst}, the model is calculated assuming black body emission described by the Planck function using the average measured temperature of the SCAL source during each scan. The emission enters the beam via the second port of the FTS, and the overall spectral response of this path is different to the path from the main telescope port, both in absolute value and spectral shape. Therefore, the instrument contribution must be subtracted from the linearised voltage before applying the extended-source calibration.

The derivation of the instrument RSRF is based on observations of the nominated SPIRE dark sky field (RA:17h40m12s, Dec:+69d00m00s (J2000)). These calibration observations were repeated on every FTS observing day. For any two spectral scans, the instrument RSRF can be isolated by comparing the measured voltage density (in V\,GHz$^{-1}$) with the models of instrument and telescope emission (in \wmhzsr). The variation in instrument temperature between the two scans means that the difference between them can be used to separate the instrument RSRF. This derivation is described in detail by \citet{fultonExpAst}, and was carried out separately for forward and reverse scans of the SMEC, and for the two epochs defined by the BSM position. The final instrument RSRF curves were determined from the mean over all available dark sky spectral pair combinations in each set in order to maximise the signal-to-noise.

\subsection{Telescope model and telescope RSRF \label{sect.tele}}

The {\it Herschel} telescope model ($M_{\rm tel}$) adopted for the FTS is constructed based on the mirror emissivity determined before launch by \citet{fischer04} and two black body emitters for the primary (M1) and secondary (M2) mirrors as 
\begin{align}\label{eq:teleModel}
M_{\rm tel} = (1 - \varepsilon_{\rm M2})\varepsilon_{\rm M1}E_{\rm corr}B(T_{\rm M1},\nu) + \varepsilon_{\rm M2}B(T_{\rm M2},\nu)\nonumber \\
\hspace{0.5cm}(\mathrm{W}\,\mathrm{m^{-2}}\,\mathrm{Hz^{-1}}\,\mathrm{sr^{-1}}),
\end{align}
where $T_{\rm M1}$ and $T_{\rm M2}$ are the mirror temperatures, $\varepsilon_{\rm M1}$ and $\varepsilon_{\rm M2}$ are the mirror emissivities, $E_{\rm corr}$ is a time dependent adjustment to $\varepsilon_{\rm M1}$ \citep{hopwood} and $B(T,\nu)$ is the Planck function.

The mirror emissivities are defined as a function of frequency following the ``dusty mirror'' sample that was fitted by \citet{fischer04} as
\begin{equation}
\varepsilon_{\rm M1} = \varepsilon_{\rm M2} = 6.1366\times10^{-5}\nu^{0.5} + 9.1063\times10^{-7}\nu, 
\end{equation}
where $\nu$ is the frequency in GHz.
Three measurements were used in the fit, all outside of the high frequency end of the SPIRE band (70, 118 and 184~$\mu$m), with a total uncertainty quoted on each measurement of $\pm$14\%. An additional point near the low frequency end of the SPIRE SLW band was also measured (at 496~$\mu$m = 604~GHz) and although it was not used in the fit, it is consistent with the fit for the dusty mirror sample. The fit is well within the errors quoted for each individual point, implying that the overall error on the fit is lower than the 14\% error on individual data points. The dusty mirror results were chosen in order to simulate the presumed conditions of the {\it Herschel} telescope in space, but there will be further systematic uncertainty on the emissivity due to deviations of the laboratory sample from the real mirror.

$T_{\rm M1}$ and $T_{\rm M2}$ are calculated from temperature timelines produced by thermistors positioned across the back of the {\it Herschel} mirrors \citep[see][]{hopwood}. For any given observation, the thermistors were read out every 512 seconds and the results averaged, both in time and over the thermistors, to provide temperature inputs for equation \ref{eq:teleModel}. Over the mission, there is a sinusoidal trend in the mirror temperatures that cycles with a time period of roughly one year. On shorter timescales, the temperatures vary depending on the schedule of pointings during each operational day. There is also an overall increase in the primary mirror temperature of $\sim$1~K over the course of the mission, which indicates a secular evolution of the telescope emissivity. This evolution could be due to a build up of dust or other time dependent factors during the mission.

In order to account for changes in the emissivity with time, a correction factor, $E_{\rm corr}$, is included in equation~\ref{eq:teleModel} to adjust the emissivity of the primary mirror. A bulk adjustment of the primary mirror emissivity by $\pm\sim$1-2\% is sufficient to bring the telescope model into agreement with the measured data \citep{hopwood}.

$E_{\rm corr}$ was derived by assuming that the telescope RSRF should remain constant with time through the mission. Observations of the dark sky field were measured on many operational days, and after subtraction of the instrument emission, these should measure only the telescope spectrum. Each instrument-subtracted spectrum was divided by its corresponding telescope model, and the deviation of these individual ratios from the average ratio was used to determine $E_{\rm corr}$ as a function of time.

In principle, the final telescope RSRF could be derived from the average of these ratios for individual observations of dark sky but in practice, it was found to be better to calculate the telescope RSRF from the {\it difference} between two scans. The differencing method takes advantage of the changing instrument and telescope temperatures in different observations to isolate the telescope RSRF from the instrument RSRF, separate model effects and increase the signal-to-noise ratio. Scan pairs covering all available dark sky scans over the mission were used in the derivation, as described by \citet{fultonExpAst}. The RSRF was calculated separately for forward and reverse scans of the SMEC, and for the two epochs defined by the BSM position. The total integration time included in each high resolution mode RSRF is 41.7~hours before OD\,1011 and 58.1~hours after OD\,1011.

However, the differencing method cannot completely separate the telescope RSRF from telescope model, due to the uncertainty in the emissivity described above and this propagates to a systematic uncertainty in the final extended calibration - see Section~\ref{sect:ext_uncert}. 

\subsection{Uranus model and point-source conversion factor \label{sect.uranus}}

The planet Uranus is used as the primary standard for SPIRE FTS point-source calibration. It has a well understood submillimetre spectrum, and is essentially point-like in the FTS beam (with an angular diameter of 3.4$^{\prime\prime}$). In addition, it has a spectrum that is virtually free of spectral features, making it a more suitable spectral calibrator than Neptune, which is used as the primary standard for the photometer \citep[see][]{bendo}. 

We adopt the {\it Herschel} ESA-4 model of the disc-averaged brightness temperature of Uranus\footnote{The ESA-4 model for Uranus is available at\\ ftp://ftp.sciops.esa.int/pub/hsc-calibration/PlanetaryModels/ESA4/.}, derived by \citet{orton} using a collision-induced absorption spectrum of a well-mixed H$_2$ and He atmosphere.  This absorption was combined with additional opacity, modelled provisionally as a mixture of H$_2$S in the deep atmosphere, that rolls off toward low pressures ($\sim$0.1~bar) to account for the observed reduction in brightness temperature at lower frequencies \citep{griffinOrton, serabyn}. The basic spectrum was calibrated against {\it Spitzer} IRS data in the mid-infrared range and against {\it Herschel} PACS photometric data in the far-infrared range. The uncertainty of this model is discussed further in Section~\ref{sect:u_model}. 

The flux-density spectrum of Uranus for the date of each observation was then calculated using the effective solid angle of the planet. In order to calculate this solid angle, an equatorial radius, $r_{\rm eq}$, of 25\,559~km and an eccentricity, $e$, of 0.21291 were used, where $e$ is defined by the equatorial and polar radii as
\begin{equation}
\label{eq:eccentricity}
e = \left[\frac{r_{\rm eq}^2 - r_{\rm pol}^2}{r_{\rm eq}^2}\right]^{1/2}.
\end{equation}
The apparent polar radius, $r_{\rm p-a}$, was calculated as
\begin{equation}
\label{eq:planetAng}
r_{\rm p-a} = r_{\rm eq}\left[1-e^2\cos^2\left(\phi\right)\right]^{1/2}\hspace{0.5cm}\mathrm{(km)},
\end{equation}
where $\phi$ is the latitude of the sub-{\it Herschel} point. The observed planetary disk was taken to have a geometric mean radius, $r_{\rm gm}$, given by
\begin{equation}
\label{eq:geometricMean}
r_{\rm gm} = \left(r_{\rm eq} r_{\rm p-a}\right)^{1/2}\hspace{0.5cm}\mathrm{(km)}.
\end{equation}
For a {\it Herschel}-planet distance of $D_{\rm H}$, the observed angular radius, $\theta_{\rm p}$, and solid angle, $\Omega_{\rm p}$, are thus
\begin{eqnarray}
\label{eq:angRad}
\theta_{\rm p} = \frac{r_{\rm gm}}{D_{\rm H}}\hspace{0.5cm}\mathrm{(rad)}
\end{eqnarray}
and
\begin{eqnarray}
\Omega_{\rm p} = \pi \theta_{\rm p}^2\hspace{0.5cm}\mathrm{(sr)}.
\end{eqnarray}
The values of $\phi$ and $D_{\rm H}$ at the time of the observation were determined from the NASA Jet Propulsion Laboratory (JPL) horizons ephemeris system \citep{giorgini}\footnote{The ephemeris can be accessed at\\ http://ssd/jpl.nasa.gov/?horizons.}.

In addition, a beam-correction factor, $K_{\rm beam}$, to account for the size of Uranus in the beam was applied, assuming a disk for the planet and a Gaussian main beam shape with FWHM values, $\theta_{\rm beam}$, from Section~\ref{sect_beam}, as
\begin{equation}
\label{eq:kbeam}
K_{\mathrm{beam}}(\theta_p,\theta_{\mathrm{beam}}) = \frac{1 - \mathrm{exp}(-x^2)}{x^2},
\end{equation}
with
\begin{equation}
x = 2\sqrt{\ln{(2)}}\left(\frac{\theta_{\rm p}}{\theta_{\rm beam}}\right).
\end{equation}

The point-source conversion factor, $C_{\rm point}$, is defined as the ratio of the Uranus model, $M_{\rm Uranus}$, to its observed spectrum,  $I_{\rm Uranus}$, calibrated as an extended source using equation~\ref{eqn_extended},
\begin{equation}
\label{eq:point}
C_{\rm point} = \frac{M_{\rm Uranus}}{I_{\rm Uranus}}\hspace{0.5cm}\left(\frac{\mathrm{Jy}}{\mathrm{W}\,\mathrm{m^{-2}}\,\mathrm{Hz^{-1}}\,\mathrm{sr^{-1}}}\right).
\end{equation}
The value of $C_{\rm point}$, as a function of frequency, was calculated separately for high and low resolution observations, and for the two epochs before and after the the change in BSM position. For off-axis detectors, observations were made with Uranus centred on each detector. The details of the Uranus observations that were used for the calibration are given in the Appendix (Table~\ref{tab_uranus}), including the measured pointing offsets determined by \citet{valtchanov}. These pointing offsets were accounted for using a Gaussian beam profile, which is a good approximation for the SSW beam where the effect of miss-pointing is most significant (see Section~\ref{sect_beam}). In order to minimise the effects of systematic additive noise described in Section~\ref{sect:tel_model}, a dark sky spectrum observed on the same day as the Uranus observation was subtracted from $I_{\rm Uranus}$ where necessary.

\section{Frequency calibration}\label{sect_freq}

\begin{figure}
\centering
\includegraphics[width=\hsize]{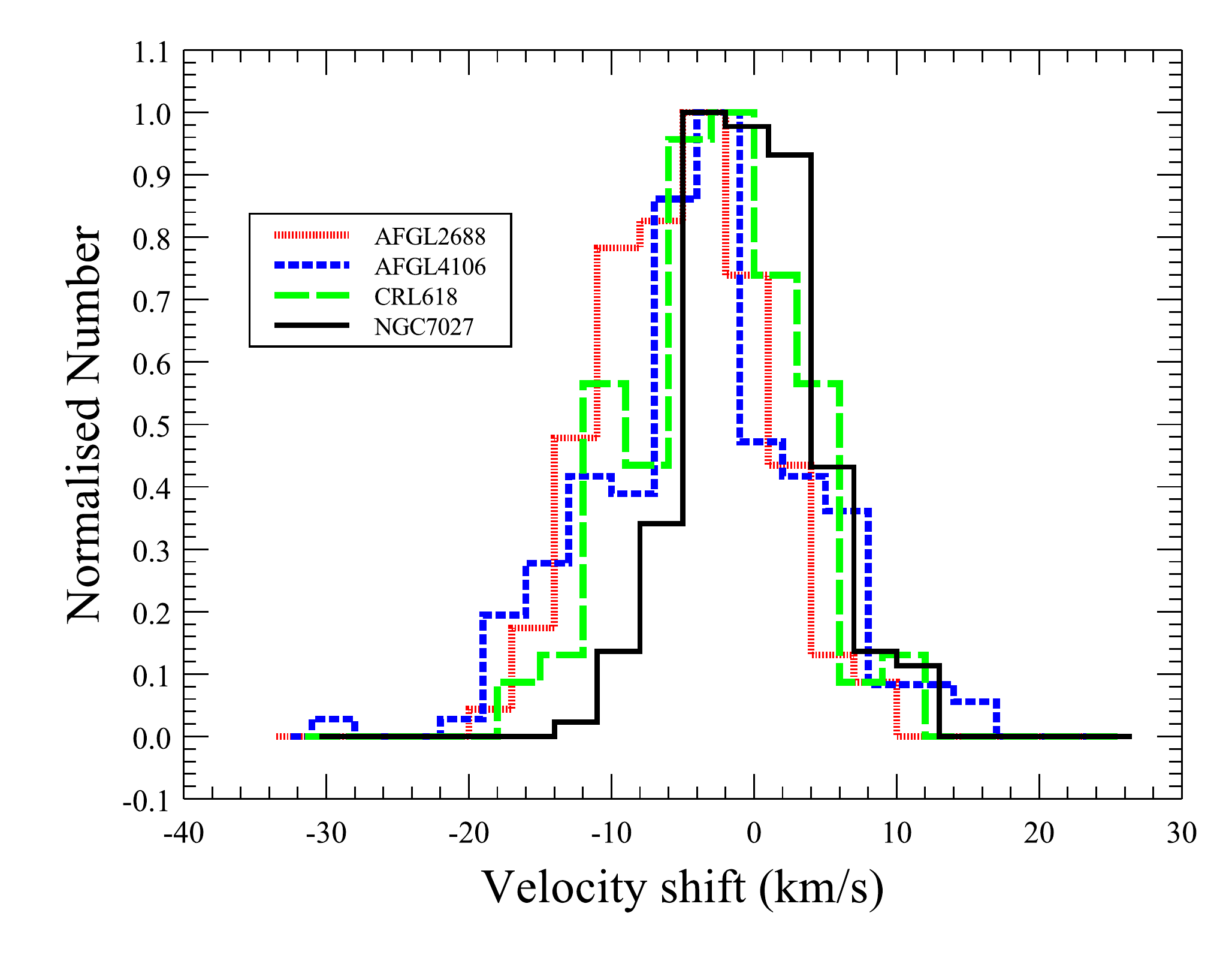}
\caption{Histogram showing the offset of the line centre from the expected source velocity for the routine calibrator sample.}
\label{freq2}
\end{figure}

The theoretical instrument line shape for any FTS is a cardinal sine function (i.e. the sinc function), unless some form of apodisation has been applied \citep[see][]{naylorTahic}. The actual measured line shape for the centre detectors has been found to be close to the classical sinc function \citep{spencer, naylor10}, which can be used to fit the observed lines accurately - see Section~\ref{sect:specLines}. The insert in the bottom plot of Fig.~\ref{interferogram} shows the shape of a typical measured spectral line.

The frequency calibration of the FTS is determined by precisely matching the measured signal with the optical path difference (OPD). For an infinitely small detector located on the optical axis, the frequency scale would be uniquely defined by the metrology system that samples the changing OPD of the moving mirror. However, the finite aperture of the detectors, their offset from the optical axis and any vignetting of the detector illumination changes the OPD \citep[due to obliquity effects;][]{spencer}, and means that the fine calibration of the frequency scale must be set by observations of known spectral lines.

In the SPIRE FTS, an optical encoder is used to determine relative changes in the position of the SMEC \citep{swinyardSpie}. This effectively counts ``steps'' in the SMEC position with a step size of 1~$\mu$m. An interpolation algorithm (Fulton et al., in preparation) is used to calculate the position of the mirror at the time of each detector sample (the detectors are read out at 80~Hz). The time constants of individual detectors and the electronics, and the phase lag associated with the readout of each detector channel, are taken into account to calculate the time of each sample. The approximate position of ZPD is set based on ground test measurements and its exact position determined during the phase correction process. 

A scale factor is also required to convert each step of the SMEC into OPD. Due to the Mach-Zehnder configuration, this step factor is approximately equal to 4, but its exact value can also be used (to first order) to correct for the obliquity effects due to the offset of each detector from the optical axis. The SMEC step factor values were determined early in the mission from three observations of the Orion Bar which were made on OD\,302, with {\it Herschel} observation identification numbers (obsids) 1342192173, 1342192174 and 1342192175. A least-squares minimisation routine was used to fit a continuum background and determine the best-fit line centres for each $^{12}$CO line in the extended calibrated Level-1 spectra from each observation. The intrinsic source velocity and the satellite velocity along the line of sight were taken into account to calculate the SMEC step factor, $f_{\mathrm{SMEC}}$, as:
\begin{equation}
f_{\mathrm{SMEC}} = 4 \frac{\nu_{\rm corr}}{\nu_{\mathrm{CO}}},
\end{equation}
where $\nu_{\rm corr}$ is the measured line centre corrected for the satellite velocity along the line of sight and the intrinsic velocity of the Orion Bar \citep[assumed to be 10~kms$^{-1}$, e.g.][]{buckle}, and $\nu_{\mathrm{CO}}$ is the rest frequency of the corresponding $^{12}$CO line. The results were averaged to obtain individual values for each detector, which are approximately 4 for the central detectors, SSWD4: 3.99923$\pm$0.00001; SLWC3: 3.99913$\pm$0.00002, and fall to $\sim$3.99 for the outer ring of detectors.

In principle, the different step factors would lead to a different frequency bin size for different detectors, but in the pipeline the interferograms are padded with zeros such that the final high resolution bin size is equal to a quarter of a resolution element (0.299~GHz) for all detectors (Fulton et al., in preparation).

In order to test the frequency calibration, the $^{12}$CO lines were also fitted in the set of observations of routine calibration line sources (AFGL2688, AFGL4106, CRL618, NGC7027), as described in Section~\ref{sect:repeatability}. The resulting distributions in the offset of the line centre from the expected source velocity, ignoring the noisy ($J$=4--3 and $J$=5--4) and blended ($J$=6--5 and $J$=10--9) lines, are shown in Fig.~\ref{freq2}. The expected line positions are reproduced with a systematic offset of $<$5~kms$^{-1}$ and a spread of $<$7~kms$^{-1}$, equivalent to approximately 1/50--1/12 of the resolution element (230--800~kms$^{-1}$ across the band). The spread is consistent with the limit with which the line centres can be determined due to the signal-to-noise ratio of the data. The systematic offset probably reflects the uncertainty in determining the expected source velocity that has been subtracted from the data, as these evolved stars all have complex line shapes with broad or asymmetric profiles \citep[e.g.][]{herpin}. The source velocities adopted for this analysis were, AFGL2688: $-35.4$~kms$^{-1}$ \citep{herpin}; AFGL4106: $-15.8$~kms$^{-1}$ \citep{josselin}; CRL618: $-25.0$~kms$^{-1}$ \citep{teyssier}; NGC7027: $+25.0$~kms$^{-1}$ \citep{teyssier}.

\section{Beam profile}\label{sect_beam}

The FTS beam profile was measured directly as a function of frequency by mapping a point-like source (Neptune) at medium spectral resolution \citep{makiwa2013}. The beam profile has a complicated dependence on frequency due to the SPIRE FTS optics and the multi-moded nature of the feedhorn coupled detectors. The feedhorns consist of conical antennas in front of a circular-section waveguide, and the diameter of the waveguide determines the cut-on frequencies of the electromagnetic modes that are propagated~\citep{chattopadhyay-ieeetrans-2003, murphy-irphys-1991}.

\citet{makiwa2013} fitted the measured Neptune data using a superposition of Hermite-Gaussian functions. They found that within the uncertainties of the measurement, only the zeroth order function (i.e. a pure Gaussian) was required for the SSW band. However, the SLW band required the first three basis functions (i.e. is not Gaussian). These fitted functions are a convenient mathematical description of the beam, rather than directly representing the electromagnetic modes propagating through the waveguides. The beam is well fitted by radially symmetric functions.

\begin{figure}
\centering
\includegraphics[width=\hsize]{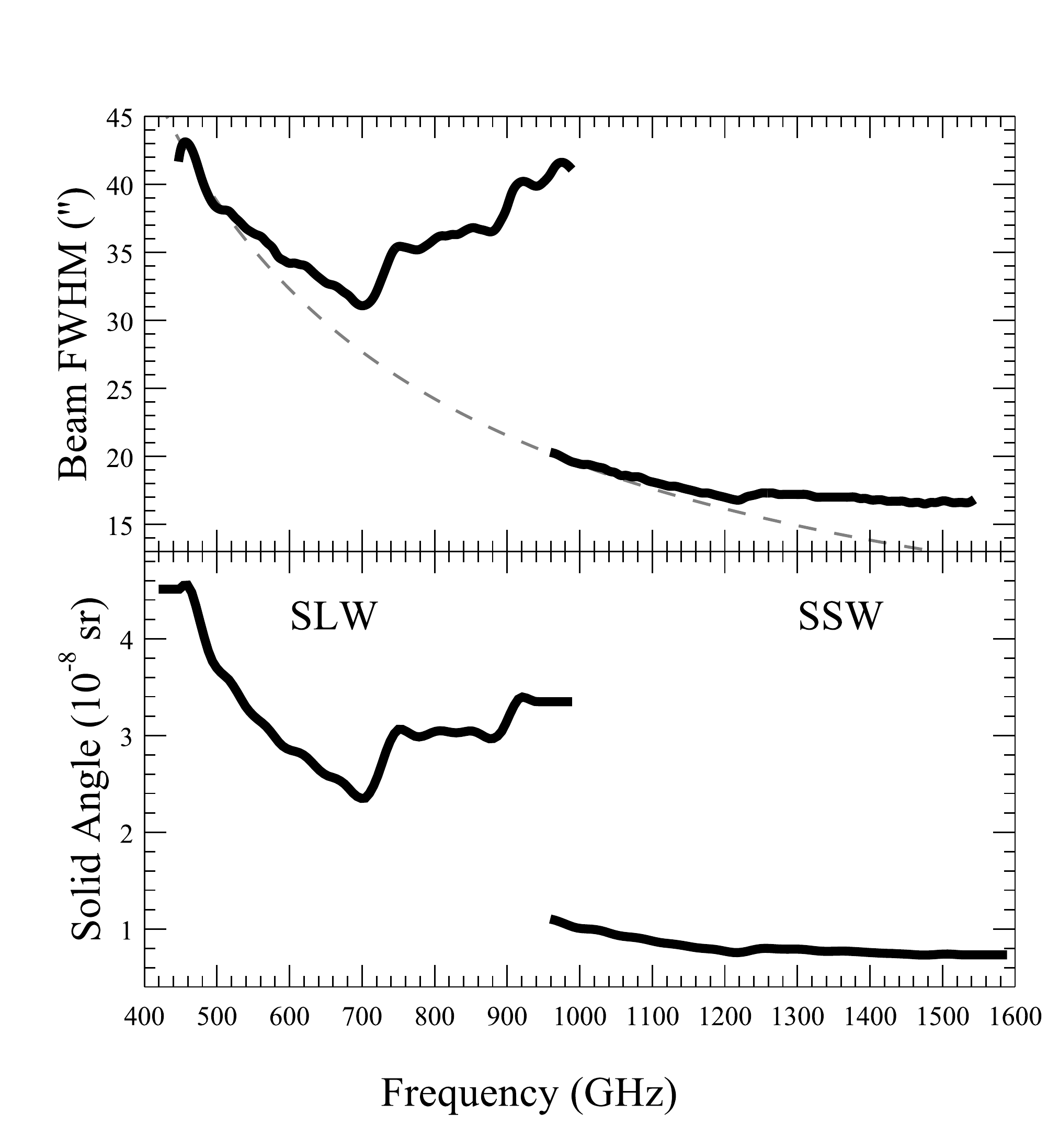}
\caption{The FWHM (top) and solid angle (bottom) of the SPIRE FTS beam profile as fitted by \citet{makiwa2013}. The grey dashed line in the top plot shows the expected FWHM from diffraction theory.}
\label{fig:fts_beams}
\end{figure}

The resulting FWHM and total solid angle of the fitted profiles are shown in Fig.~\ref{fig:fts_beams}, which also includes the expected frequency dependence for diffraction alone (calculated as the FWHM of an Airy pattern with effective mirror diameter of 3.287 m). The beam size matches the diffraction limit only at the low-frequency end of each band, where the waveguide is single moded. As further modes propagate, their superposition leads to beam sizes larger than expected from diffraction theory.

The beam profile shapes are included with the calibration data attached to each SPIRE FTS observation in the {\it Herschel} Science Archive, and can be used to correct for the frequency dependent source-beam coupling if a good model of the source spatial distribution is available \citep{wu}.

\section{Accuracy and repeatability}\label{sect_accuracy}

\subsection{Sensitivity achieved}\label{sect_sens}

There are three main sources of uncertainty associated with processed FTS spectra -- random noise, which can be reduced by integrating over a greater number of scans; broad systematic features affecting the continuum shape (on a scale much wider than the spectral resolution element); and small scale systematic effects associated with the observed signal (fringing, and ringing in the instrumental line shape wings, which both increase with source strength). For very bright point sources, pointing jitter also contributes to the noise in the spectra.

We can deduce the expected random noise and uncertainty contributions from the various parameters used in the conversion of the observed voltage to intensity and flux density using equations~\ref{eqn_extended} and~\ref{eqn_point}.  If we assume that the transformation from the time domain interferogram to the frequency domain spectrum preserves the stochastic and systematic uncertainties, and that the process adds no extra uncertainty term, then we can write the estimated noise on the extended calibrated spectrum as
\begin{align}\label{eqn:ext_error}
\delta I_{\rm ext} = \frac{V_{\rm obs}}{R_{\rm tel}} \sqrt{ \left(\frac{\delta V_{\rm obs}}{V_{\rm obs}}\right)^{2} + \left(\frac{\delta R_{\rm tel}}{R_{\rm tel}}\right)^{2}} \nonumber
  \mspace{100mu}
  \notag\\
\hspace*{0.5cm}+M_{\rm inst}\frac{R_{\rm inst}}{R_{\rm tel}}\sqrt{\left(\frac{\delta R_{\rm inst}}{R_{\rm inst}}\right)^{2} + \left(\frac{\delta R_{\rm tel}}{R_{\rm tel}}\right)^{2}}.
\end{align}
The first term takes account of the random error on the measured signal divided by the telescope RSRF. The second term takes account of the instrument model and must be added linearly as it has a systematic effect on $I_{\rm ext}$. Note that the contribution from the instrument model is completely insignificant where the instrument emission (at $\sim$4.5~K) becomes vanishingly small in the SSW band. Equation~\ref{eqn:ext_error} assumes that there is no contribution to ${\delta}I_{\rm ext}$ from the instrument and telescope models themselves and the additional systematic uncertainties due to these models are discussed further in Section~\ref{sect:tel_model}.

Conversion from the extended calibration to point-source flux density, via equation~\ref{eqn_point}, and the derivation of $C_{\rm point}$ in equation~\ref{eq:point}, inserts a further uncertainty term to reflect the stochastic error on the measurement of the Uranus spectrum,
\begin{equation}\label{eqn:point_error}
\delta F_{\rm point} = F_{\rm point}\sqrt{\left(\frac{\delta I_{\rm obs}}{I_{\rm obs}}\right)^2 + \left(\frac{\delta I_{\rm Uranus}}{I_{\rm Uranus}}\right)^2}.
\end{equation}
The additional systematic uncertainty on the Uranus model is discussed further in Section~\ref{sect:u_model}.

\begin{figure}
\centering
\includegraphics[width=\hsize]{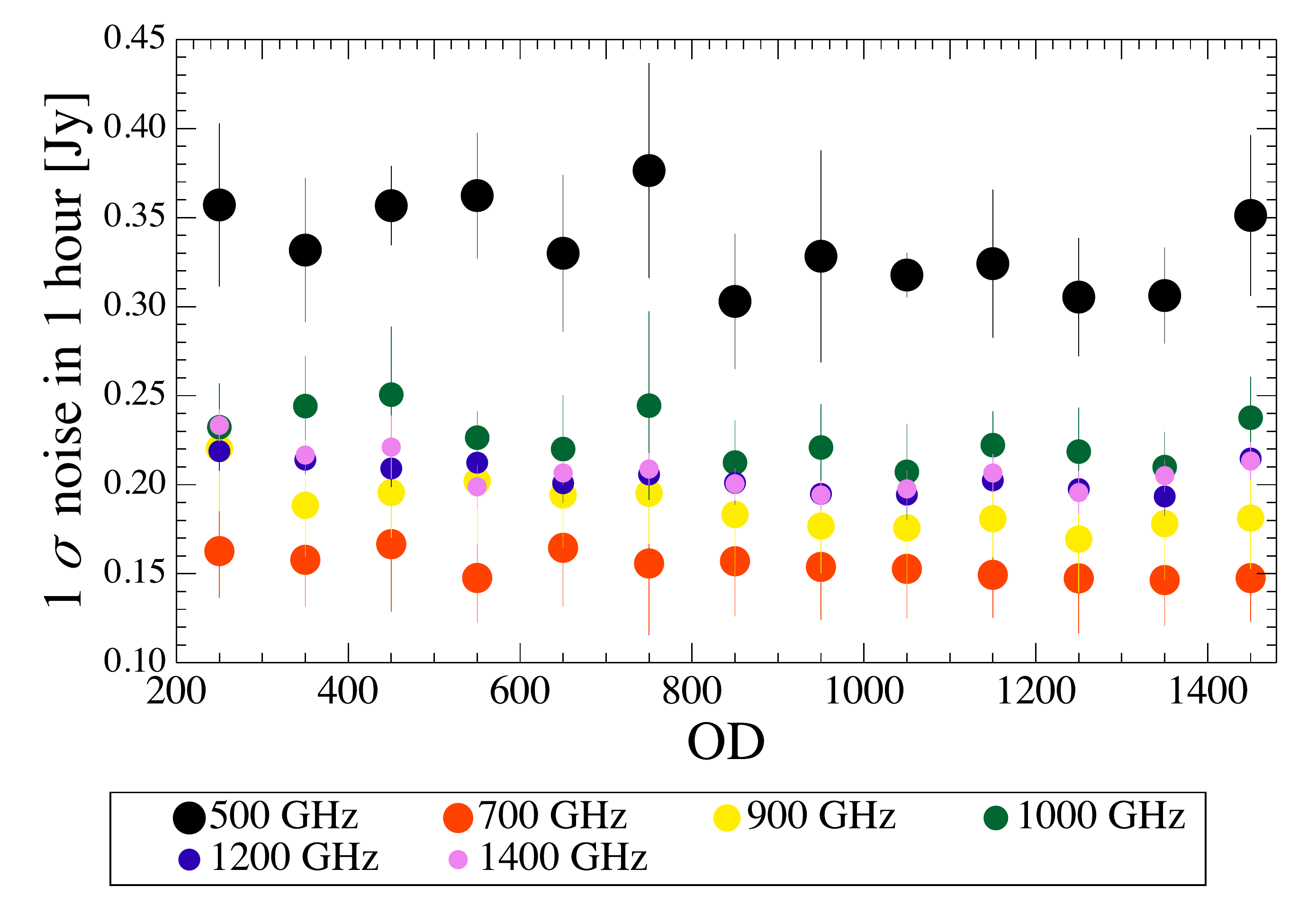}
\caption{Noise within the 1.2 GHz instrument resolution element width estimated for Level-2 dark sky observations as a function of operational day (OD). The noise has been rebinned in 200~GHz steps centred at the frequencies shown in the legend. The noise shows no significant systematic trends for any frequency over the course of the mission.}
\label{fig:noiseWithOd}
\end{figure}

\begin{figure}
\centering
\includegraphics[width=\hsize]{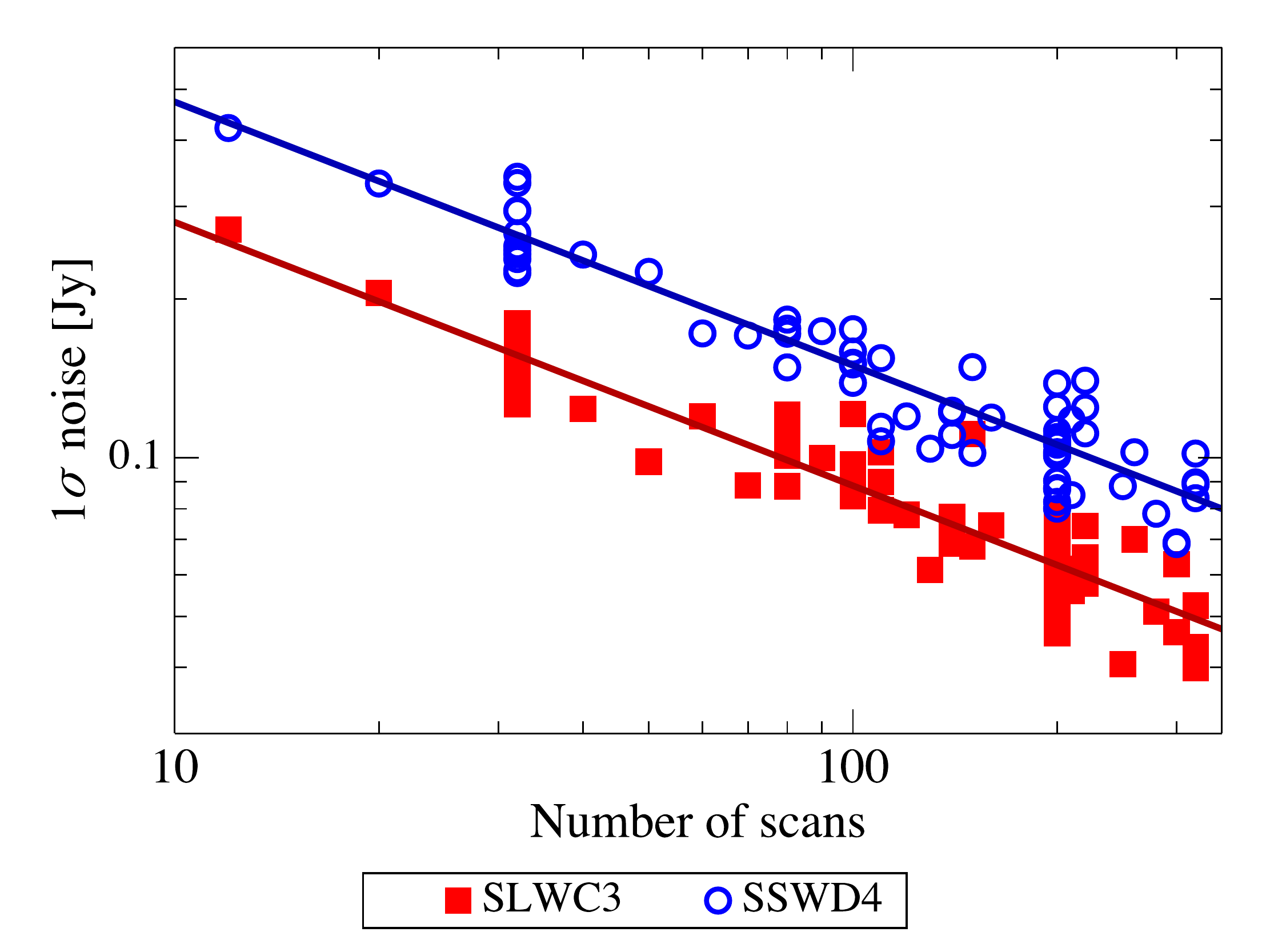}
\caption{Average noise estimated for Level-2 dark sky observations as a function of number of scans ($n$). The symbols (red squares: SLWC3 and blue circles: SSWD4) are the noise estimates and the corresponding straight lines are fitted functions of the form $1/\sqrt{n}$, showing the noise integrates down as expected up to the longest dark sky observations taken.}
\label{fig:noiseWithScans}
\end{figure}

For any processed observation, the standard pipeline attaches the standard error on the mean of the repeated scans in an ``error" column. However, these values represent only the random noise component. In order to take account of all the terms from equations~\ref{eqn:ext_error} and \ref{eqn:point_error}, the noise was measured directly from the final spectrum as the standard deviation in 50~GHz frequency bins. Analysis of observations of dark sky with more than 20 repetitions shows that noise levels have remained consistent throughout the mission, across the whole frequency band (see Fig.~\ref{fig:noiseWithOd}) and that noise levels roughly integrate down as expected with increasing number of scans, up to the longest dark sky observations taken (see Fig.~\ref{fig:noiseWithScans}).

\begin{figure}
\centering
\includegraphics[width=0.65\hsize]{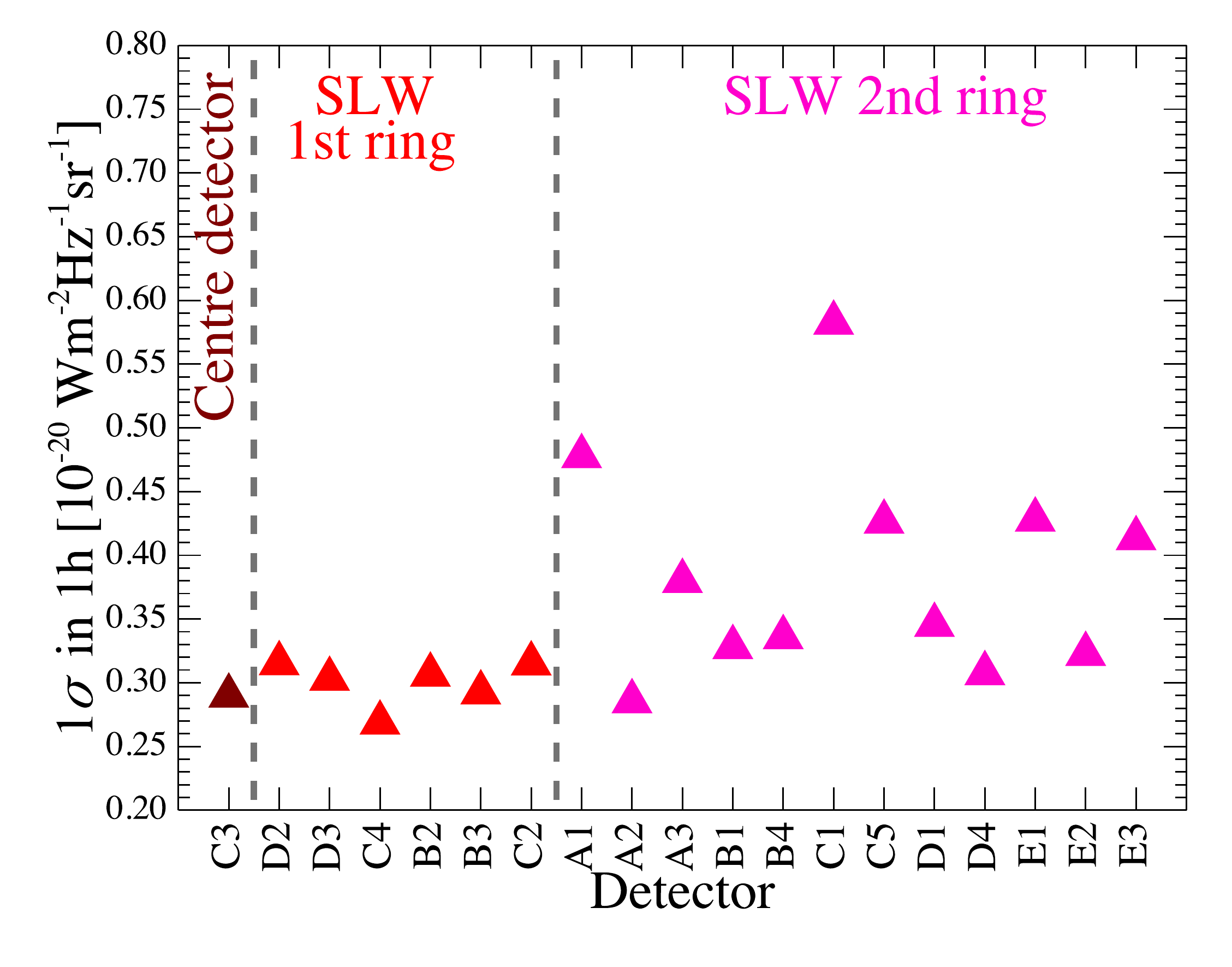}
\includegraphics[width=\hsize]{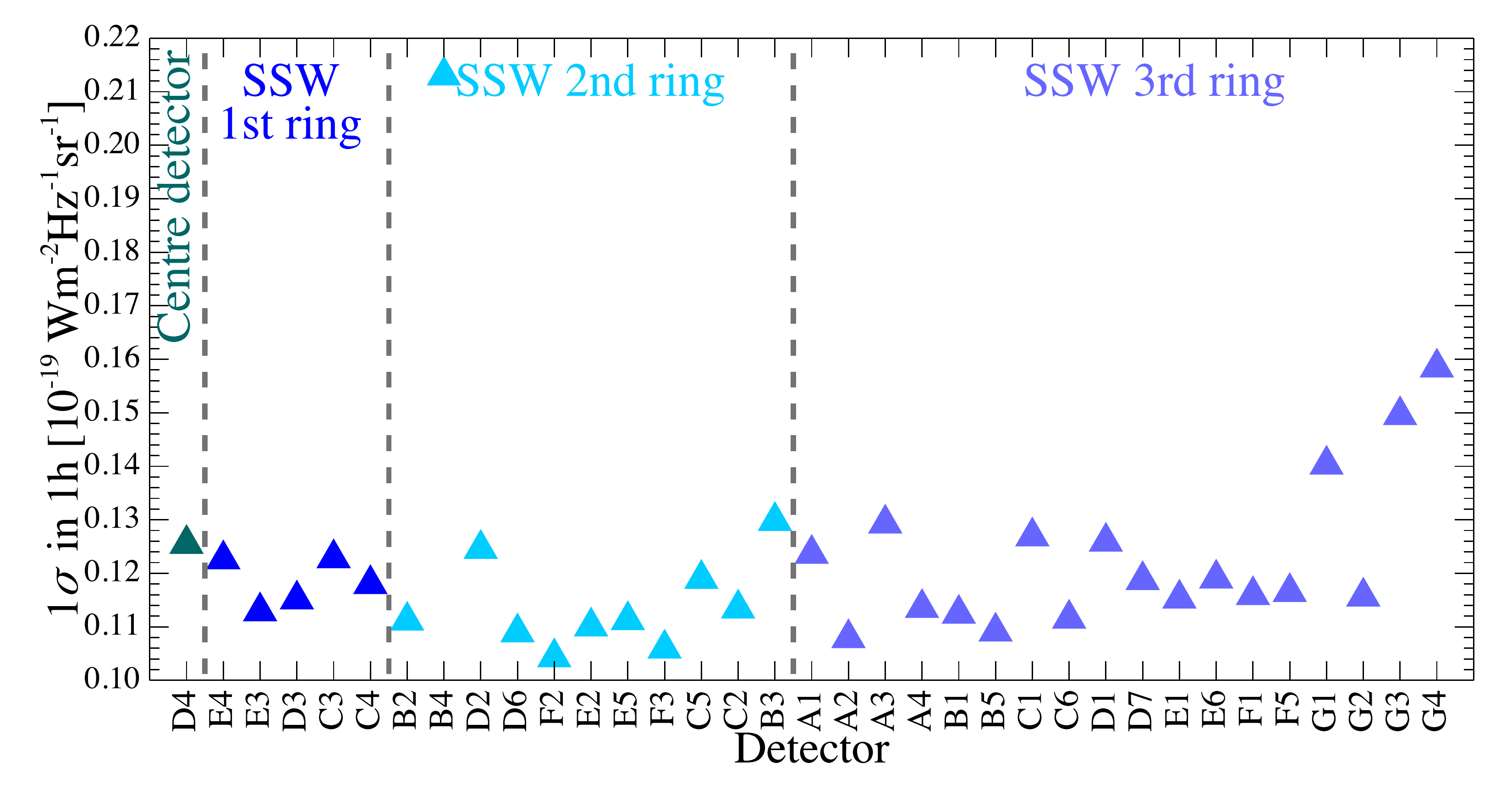}
\caption{Average sensitivity (taken across the entire band) for all detectors for extended-source calibrated data. The point-source-calibrated sensitivity for the centre detectors is shown in Fig.~\ref{fig:noiseCheck} as a function of frequency. The sensitivity is consistent across the unvignetted detectors, with greater scatter for the outer (vignetted) rings. Note the SLW sensitivity is approximately an order of magnitude better than for SSW ($1\times10^{-19}$ compared to $1\times10^{-20}$).}
\label{fig:sensitivity}
\end{figure}

The noise results from the dark sky, and a few sources that show no significant spectra features (Uranus and Ceres) provide a realistic estimate of the minimum detectable signal (i.e. the sensitivity) as a function of frequency. When the data are processed using the extended-source calibration, the spectrum is calculated in intensity units (\wmhzsr), and as the beam size is smaller for SSW, the noise levels are far higher in SSW, compared to SLW,  as shown in  Fig.~\ref{fig:sensitivity}. The values are consistent for the unvignetted detectors, but have a higher scatter for the outer detectors, which are vignetted. 

The point-source sensitivity, calculated with respect to the in-beam flux density, results in similar noise levels in the two bands, with the levels rising towards the edges of the band. The yellow line in Fig.~\ref{fig:noiseCheck} shows the point-source sensitivity in Jy as a function of frequency for the centre detectors, and is discussed further in Section~\ref{sect:exp_noise}. 

\subsection{Comparison with expected noise}\label{sect:exp_noise}

In this section, we assess whether the measured noise is commensurate with what is expected by modelling the detector and photon noise, following the method presented in \citet{sensitivityNote}. The power falling onto the detectors from the telescope, $Q_{\rm tel}$, can be estimated using the transmission curves, generated before flight, from measurements of the individual components in the optical chain as
\begin{equation}
Q_{\rm tel} = \eta_{\rm inst}\int\limits^{\nu_u}_{\nu_l}\!t_{\rm band}(\nu)A\Omega(\nu)M_{\rm tel}(\nu)\,{\rm d}\nu\hspace{0.5cm}(\mathrm{W}),
\end{equation}
where $A$ is the area of the telescope, $\Omega(\nu)$ is the solid angle of the beam for an individual detector (see Fig.~\ref{fig:fts_beams}), $t_{\rm band}(\nu)$ is the overall transmission curve for the band in question normalised to its peak value and $\eta_{\rm inst}$ is the product of the overall coupling efficiency and peak transmission between the instrument and the source. We assume that for extended sources, the coupling efficiency is accounted for in the calculation of $A\Omega(\nu)$ and so $\eta_{\rm inst}$ represents only the peak transmission efficiency of the instrument.  For point sources (see below) this is not necessarily true and an additional factor is required.

$Q_{\rm tel}$ affects the overall noise equivalent power (NEP) in two ways: by directly contributing photon noise ($\mathit{NEP}_{\rm phot}$), and by determining the bolometer operating point and thus affecting the inherent detector NEP ($\mathit{NEP}_{\rm det}$). $\mathit{NEP}_{\rm phot}$ can be estimated as
\begin{equation}
\mathit{NEP}_{\rm phot} = (2Q_{\rm tel} h\nu)^{1/2}\hspace{0.5cm}(\mathrm{W\,Hz^{1/2}}).
\end{equation}

Calculation of $\mathit{NEP}_{\rm det}$ is based on the bolometer model described by \citet{mather} and \citet{sudiwala}, with bolometer parameters measured in the lab \citep{hien2004} and updated during the mission. Taking the estimated model value of $\mathit{NEP}_{\rm det}$ and the estimated $\mathit{NEP}_{\rm phot}$, we can calculate the minimum detectable flux density, or Noise Equivalent Flux Density ($\mathit{NEFD}$),
\begin{equation}
\label{equ:nefd}
\mathit{NEFD}(\nu) =  \frac{10^{26}2^{1/2}\sqrt{\mathit{NEP}^2_{\rm det}+\mathit{NEP}^2_{\rm phot}}}{\eta_{\rm point} \eta_{\rm cossq}At_{\rm band}(\nu)\Delta\nu}\hspace{0.5cm}(\mathrm{Jy\,Hz^{1/2}}),
\end{equation}
where $\eta_{\rm point}$ expresses any difference in coupling efficiency between a point source and an extended source and $\eta_{\rm cossq}$ the loss in efficiency due to the Fourier transform nature of the detection. Depending on the definition of NEP, this can be taken to be either 0.5 or 1/$\sqrt{8}$ \citep{treffers}. Here, we define NEP as the equivalent power that gives a signal-to-noise ratio of 1 in a 1\,s integration time and therefore include a factor of $\sqrt{2}$ in equation~\ref{equ:nefd} (see also equation~\ref{equ:delta_s}, below), thus we adopt $\eta_{\rm cossq}=0.5$. $\Delta\nu$ is the width of the spectral resolution element - assumed here to be 1.2 GHz (i.e. for the high resolution mode of the FTS). The sensitivity in a one hour observation is then given as:
\begin{equation}
\label{equ:delta_s}
\Delta F_{\rm point}(\nu) = \mathit{NEFD}(\nu)/(2^{1/2}3600^{1/2})\hspace{0.5cm}(\mathrm{Jy}).
\end{equation}

\begin{figure}
\centering
\includegraphics[width=\hsize]{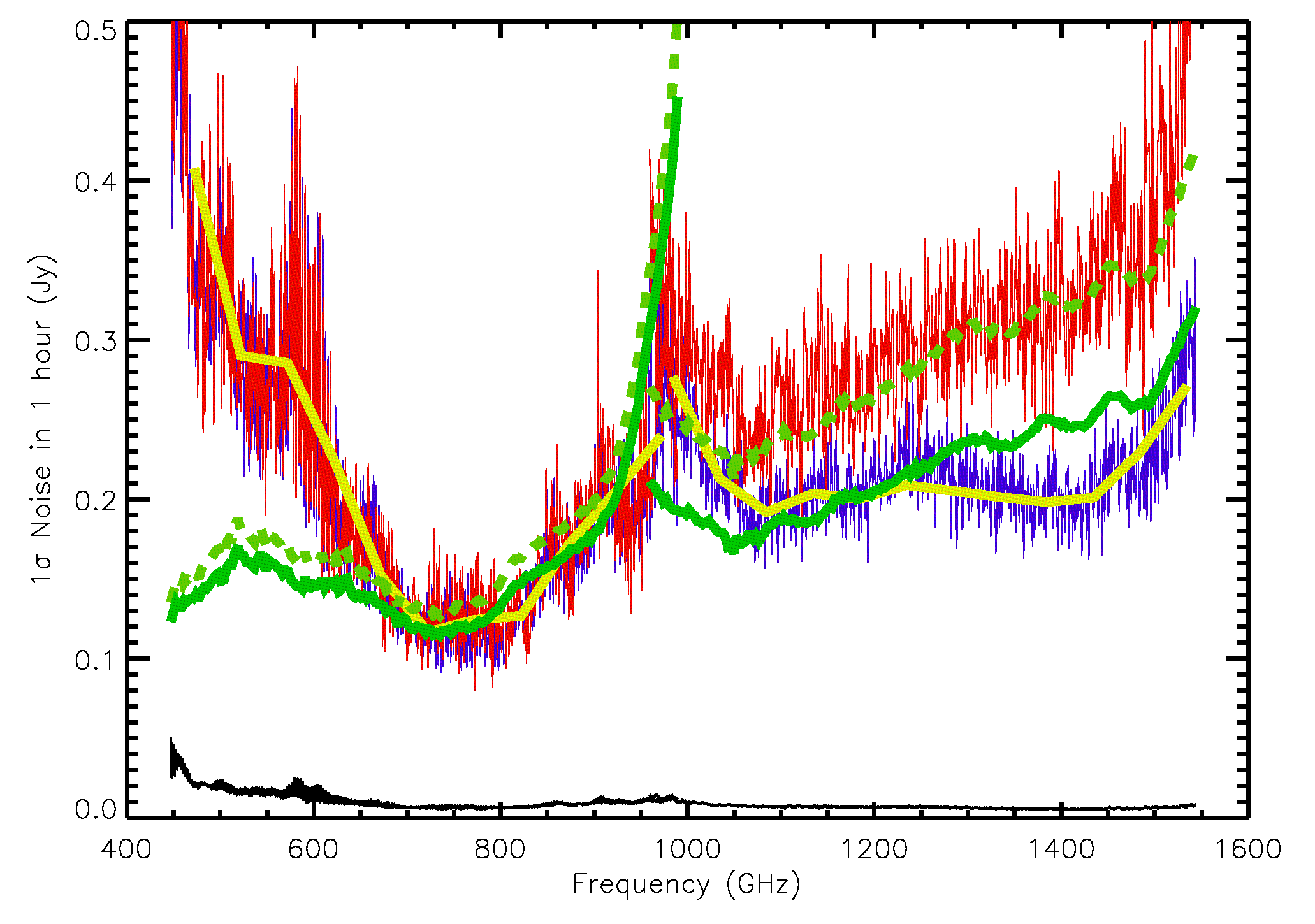}
\caption{The yellow solid line shows the overall measured sensitivity based on the analysis of multiple dark sky observations and expressed as the 1$\sigma$ minimum detectable flux density (Jy) in a one hour integration. The red and blue lines represent the measured random noise on the spectrum of Eta Carina and dark sky respectively, both referenced to a one hour integration. The green dashed and solid lines are the expected sensitivity level for Eta Carina and dark sky respectively, derived from equation~\ref{equ:delta_s}. The black line shows the uncertainty associated with the instrument and telescope RSRFs.}
\label{fig:noiseCheck}
\end{figure}

In order to calculate the expected sensitivity, we used an $\eta_{\rm inst}$ value of 0.45 for SLW and 0.35 for SSW, and an $\eta_{\rm point}$ value of 0.7 for SLW and 0.65 for SSW. These values are in line with expectations from pre-flight measurements. In equation~\ref{equ:delta_s}, we include a factor of $\sqrt{2}$ in the numerator to account for the definition of NEP used, as mentioned above. Fig.~\ref{fig:noiseCheck} compares the measured random noise (as given by the ``error'' column) on one dark sky observation (obsid 1342197456), and the bright star Eta Carina (obsid 1342228700) with the estimated expected sensitivity calculated above. The uncertainty associated with the telescope and instrument RSRFs (the second term in equation~\ref{eqn:ext_error}) is also plotted in Fig.~\ref{fig:noiseCheck}, showing that it is significantly lower than the measured and expected noise. 

These results show that the measured and expected sensitivity are in good overall agreement.  The differences in the detailed shape of the curves are due to the actual transmission curves of the instrument deviating from the product of the individual filter transmissions, and some possible loss of point-source efficiency towards the low-frequency end of the SLW detectors \citep[see][]{chattopadhyay-ieeetrans-2003, wu}.

The final measured sensitivity (the yellow curve plotted in Fig.~\ref{fig:noiseCheck}) represents a significant improvement with respect to previous versions of the data processing pipeline, in particular due to improvements in the signal-to-noise ratio of the RSRFs which were implemented for HIPE Version 11 \citep[see][]{fultonExpAst}.

\subsection{Point-source calibration accuracy: systematic uncertainty in the Uranus model}\label{sect:u_model}

\begin{figure}
\centering
\includegraphics[trim = 0mm 0mm 15mm 170mm, clip, width=\hsize]{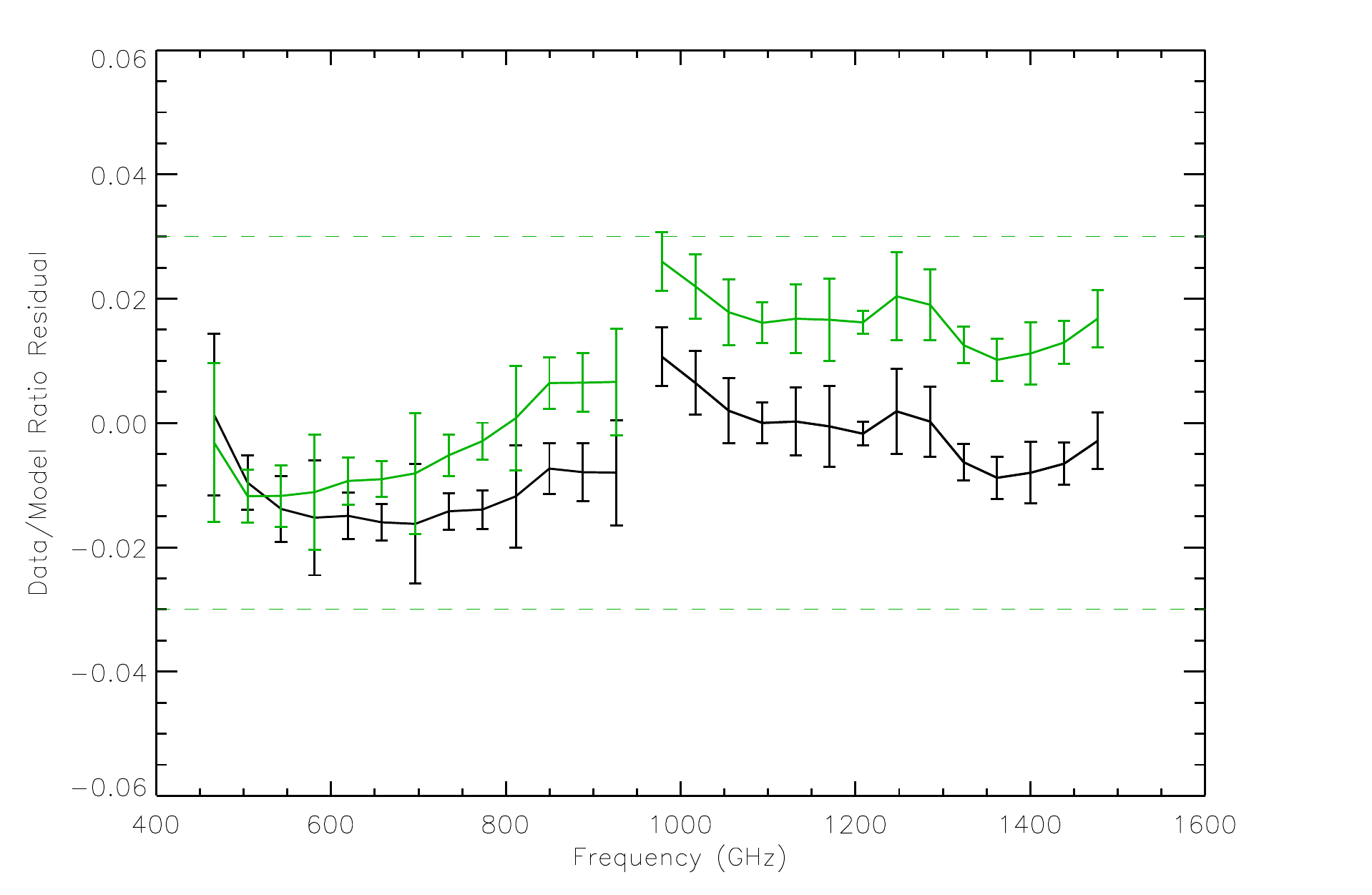}
\includegraphics[trim = 0mm 0mm 15mm 170mm, clip, width=\hsize]{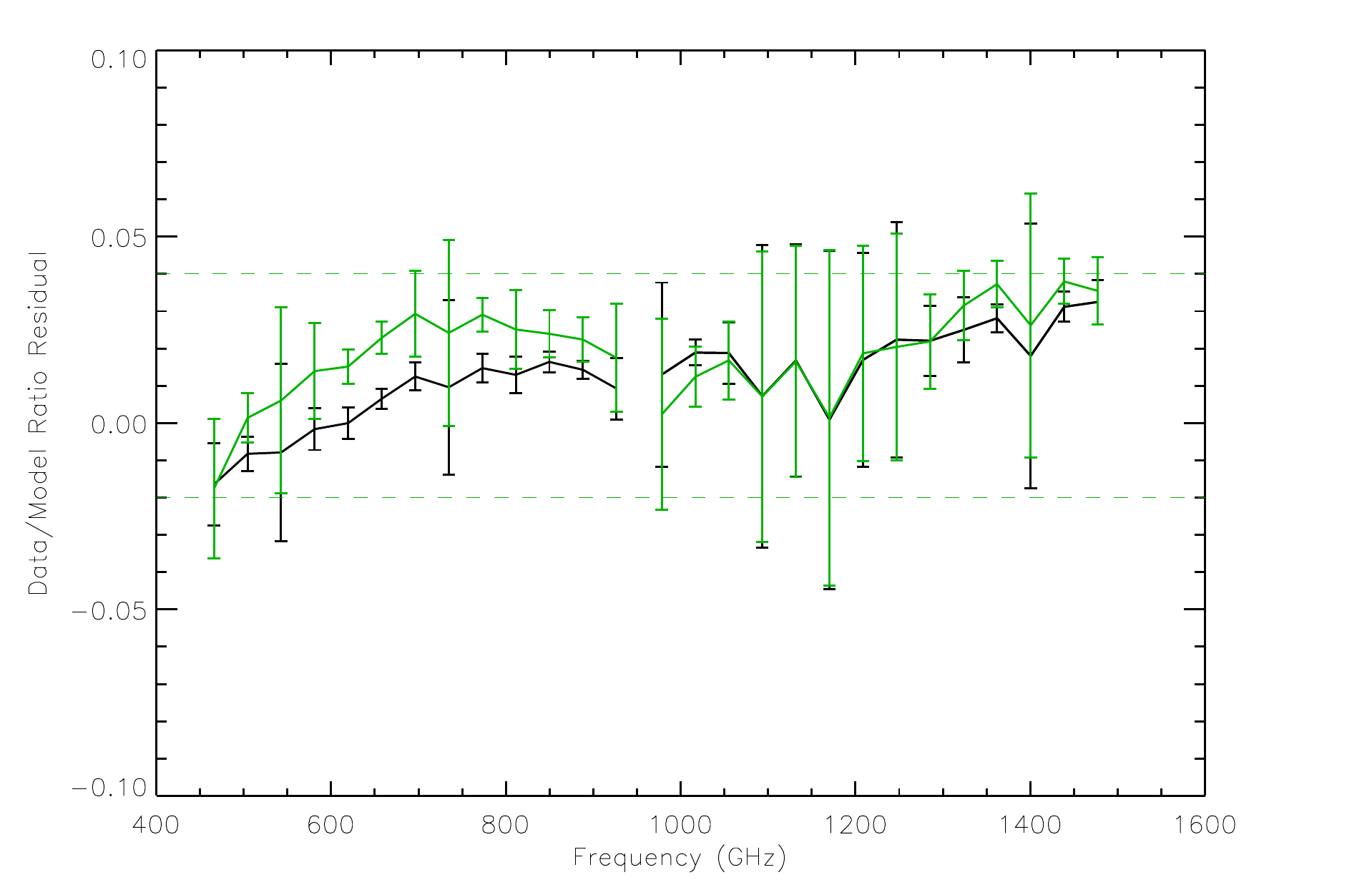}
\caption{The top plot shows the Neptune/Uranus data-to-model ratio (from equation~\ref{eqn_nep_ur_rat}) using the two different Uranus models, ESA-4 (black) and ESA-5 (green). The bottom plot shows the Mars/Neptune data-to-model ratio in black and the Mars/Uranus data-to-model ratio in green. The dashed horizontal lines represent an arbitrary $\pm$3\% uncertainty limit centred on 0\% for the top plot and +1\% for the bottom plot.}
\label{fig:models}
\end{figure}

The absolute calibration of point-source spectra relies on the model of Uranus used to derive the $C_{\rm point}$ parameter in equation~\ref{eq:point}. This section describes tests of the model accuracy in the SPIRE frequency range. Two other planetary sources commonly used for calibration in the IR and sub-mm, Neptune and Mars, were used for a relative comparison with Uranus.

Neptune was observed on a number of occasions during the {\it Herschel} mission, including an observation made on 2010-06-09 (OD\,392; obsid 1342198429) that has been shown to be well pointed \citep{valtchanov}. A similarly well pointed observation of Uranus, made on 2010-05-31 (OD\,383; obsid 1342197472), was used for the comparison.

The model of Neptune was calculated from the disk-averaged brightness temperature spectrum based on the ESA-4 version of the planetary atmosphere model first published by \citet{moreno}\footnote{The ESA-4 model for Neptune is available at\\ ftp://ftp.sciops.esa.int/pub/hsc-calibration/PlanetaryModels/ESA4/.}.
To test the relative accuracy of the models, the ratio of the point-source-calibrated spectrum of Neptune to the spectrum of Uranus was calculated. This ratio was divided by the ratio of their respective models to give $r_{\rm test}$, which is effectively independent of the Uranus model used for the FTS calibration and gives a measure of the absolute calibration accuracy of the two model spectra. $r_{\rm test}$ is shown in Fig.~\ref{fig:models}, defined as,
\begin{equation}\label{eqn_nep_ur_rat}
r_{\rm test} = \frac{{F_{\rm Neptune}}/{F_{\rm Uranus}}}{{M_{\rm Neptune}}/{M_{\rm Uranus}}}-1.
\end{equation}

Two Uranus models were tested, as the modelled spectrum is sensitive to the precise choice of the thermal profile. The ESA-4 model (as used in the pipeline; see Section~\ref{sect.uranus}) uses the thermal profile that formed the initial basis of the analysis by \citet{feuchtgruber}. As an alternative approach, the mid-infrared {\it Spitzer} SL1 spectrum was recalibrated to account for the true angular size of Uranus \citep{orton}, and this updated model is termed ESA-5. The new model changes the overall brightness temperature in the SPIRE band by $\pm$2\%. Discounting strong spectral features, Fig.~\ref{fig:models} shows that the Neptune/Uranus ratio lies within a range of $\pm$2\% no matter which model is used. This ratio is in good agreement with the Neptune/Uranus comparison carried out for the SPIRE photometer by \citet{bendo}.
  
Mars has a well constrained model continuum due to a) its very thin atmosphere and b) the large number of orbital satellites that have measured its surface emissivity and temperature. However, Mars is very bright and the SPIRE FTS could only observe it in bright-source mode. A well pointed observation of Mars made on 2012-06-30 (OD\,1144) between 21:46 and 22:13 (obsid 1342247563). The calibration of bright-source observations is described in \citet{lu}. 

A high spectral resolution model of Mars was constructed based on the thermophysical model of \citet{rudy}, updated to use the thermal inertia and albedo maps (0.125 degree resolution) derived from the {\it Mars Global Surveyor} Thermal Emission Spectrometer (5.1--150~$\mu$m) observations \citep{putzig}. These new maps were binned to 1~degree resolution. A dielectric constant of 2.25 was used for latitudes between 60~degrees South and 60~degrees North. As in the original Rudy model, surface absorption was ignored in the polar regions and a dielectric constant of 1.5 in the CO$_2$ frost layer was assumed. Disk-averaged brightness temperatures were computed over the SPIRE frequency range and converted to flux densities for the time of the SPIRE FTS observation.

The lower plot in Fig.~\ref{fig:models} shows the equivalent value of $r_{\rm test}$ calculated for Mars versus Uranus, and Mars versus Neptune. There is a consistent variation between both outer planet spectral ratios and the Martian spectral ratio, indicating that the bright-source processing may leave some residual non-linearity. \citet{lu} found that the bright and nominal modes agree within $\pm$2\% and the observed offset is within this value. The accuracy of the Mars model is discussed in \citet{mellon}, who quote a continuum measurement error of 2.7\% and a range in the thermal model of 1.4\%. They also include errors due to interpolation across gaps in the coverage of their maps of the Martian surface, but these should not be important for our comparison because we calculate the integrated model over the entire visible hemisphere. Fig.~\ref{fig:models} shows that both the Uranus and Neptune models are consistent with the Martian model to better than 3\%.

The uncertainty in the Uranus spectral model is less due to measurement errors from {\it Spitzer} than systematic ones. The {\it Spitzer} measurement errors translate into an absolute temperature uncertainty of only $\pm$0.2~K and, in turn, propagate into a radiance uncertainty of only $\pm$0.4\% in the SPIRE bands. One source of systematic uncertainty is the assumed He vs. H$_2$ molar fractions in the models. For example, the $\pm$0.033 uncertainty associated with this ratio from the {\it Voyager} radio-occultation and infrared experiment \citep{conrath} translates into a radiance uncertainty on the order of $\pm$0.5--1.0\% over the SPIRE spectral bands. Other sources of systematic uncertainty remain to be evaluated.

Overall, the parameter driven model radiance range is largely encompassed by the variations between the ESA-4 and ESA-5 incarnations shown in Fig.~\ref{fig:models}, with a maximum excursion of $\pm$2\%. \citet[][]{bendo} quote a model range of $\pm$4\% for the Neptune ESA-4 model used for the photometer.  Given the consistency of the Neptune model to both the Uranus and Mars models, shown here, we can state that both the SPIRE FTS and photometer absolute point-source calibration accuracy can actually be quoted as $\pm$3\%. As this is a systematic uncertainty it should be combined linearly with any randomly induced uncertainty arising from uncorrected pointing errors, photon noise, etc.

\subsection{Extended source calibration accuracy}\label{sect:ext_uncert}

In principle the accuracy of the extended source calibration could be tested using a SPIRE FTS observation of a fully extended source of known surface brightness. Unfortunately, astronomical sources with flat spatial extent and well measured surface brightness across the SPIRE spectral range are not available. Therefore, the calibration accuracy can only be checked using the telescope model described in Section~\ref{sect.tele}, and is limited by our knowledge of the absolute emissivity of the telescope. 

A different method of testing the extended source calibration is described in \citet{wu}, where the problem of semi-extended source calibration is extensively discussed. In that method, the spectrum of Uranus processed using the extended source calibration is divided by its model spectrum, giving a measure of the beam size of the instrument on the sky divided by the ``coupling efficiency'' between the beam and a point source. This is equivalent to the inverse of $C_{\rm point}$ as given in equation~\ref{eq:point} (canceling the units to give a value in steradians) - i.e. a direct measure of the effective beam size,
\begin{equation}\label{eq:coupling}
\Omega_{\rm eff} = \frac{I_{\rm Uranus}}{M_{\rm Uranus}} = \frac{\Omega_{\rm beam}}{\eta_{\rm coupling}}\hspace{0.5cm}(\mathrm{sr}),
\end{equation}
where $\Omega_{\rm beam}$ is the beam as measured using a point source (see Section~\ref{sect_beam}) and $\eta_{\rm coupling}$ is a the coupling efficiency to a point source, i.e. a term needed to account for any diffraction losses and variation in the coupling to the detectors between a source that fills the field of view and a point source. 

\begin{figure}
\centering
\includegraphics[width=\hsize]{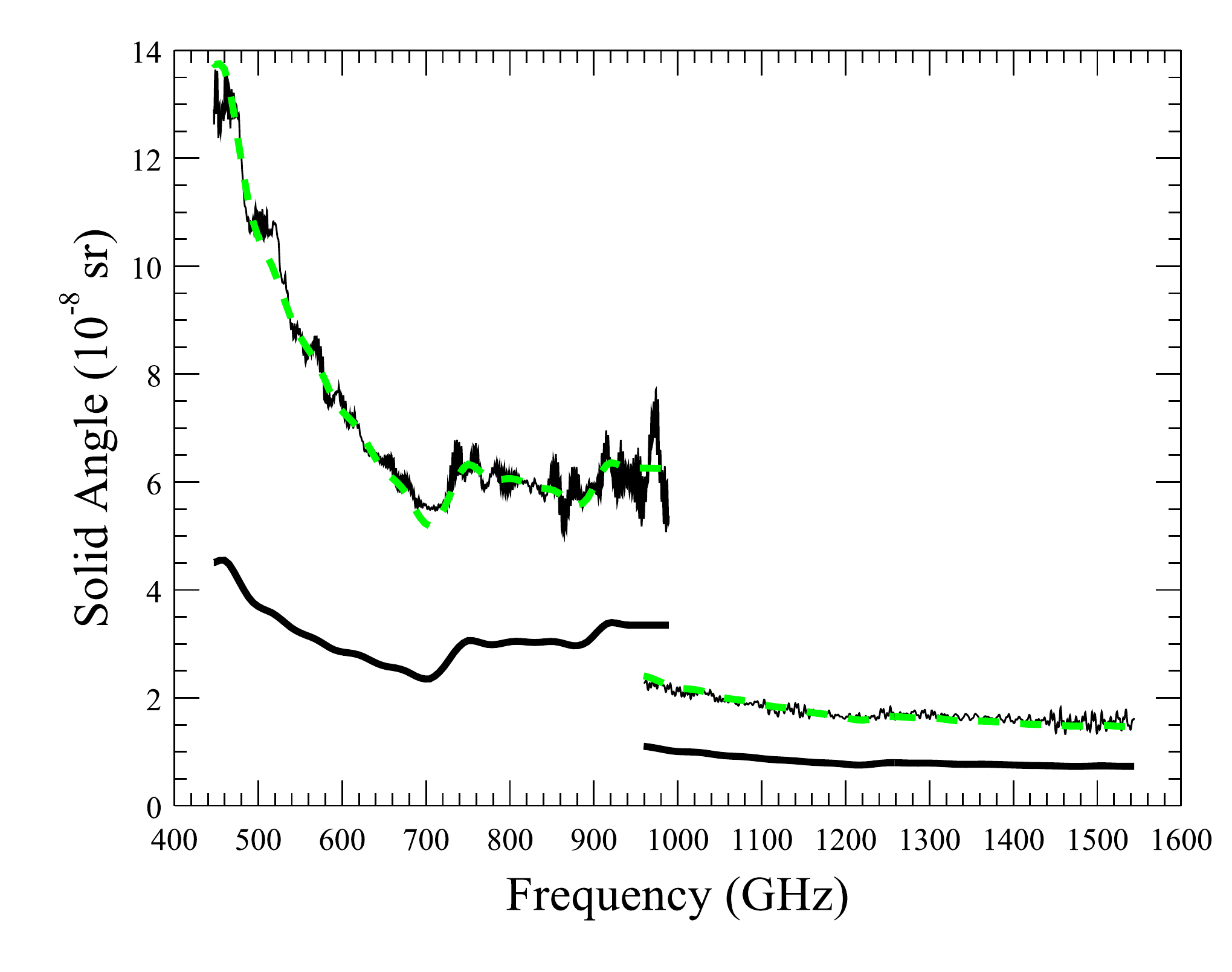}
\caption{Effective beam size in steradians ($\Omega_{\rm eff}$) calculated from equation~\ref{eq:coupling} (upper black curve). Also shown are the measured beam size for a point source (lower black line) and the beam size adjusted using a simple diffraction model and an estimation of the coupling efficiency difference between point and extended sources ($\Omega_{\rm eff}^\prime$ - green dashed line).}
\label{fig:coupling}
\end{figure}

There are two distinct contributions to $\eta_{\rm coupling}$: the first is the ``diffraction loss'' due to the coupling between the acceptance beam of the instrument and the diffraction pattern of a point source from the telescope. The second contribution is the effective absorption efficiency of the electromagnetic modes within the detector - the so-called ``feedhorn coupling efficiency''. In \citet{wu}, a simple model of diffraction losses for a point source and the feedhorn coupling efficiency for SLW \citep[as directly measured before launch by][]{chattopadhyay-ieeetrans-2003} is presented in comparison with the derived efficiency from equation~\ref{eq:coupling}, and found to be in good agreement. The diffraction model combined with a linear fit to the \citet{chattopadhyay-ieeetrans-2003} values can be used to derive $\Omega_{\rm eff}^\prime$, the expected effective beam size when observing a fully extended source. This is plotted in Fig.~\ref{fig:coupling} together with $\Omega_{\rm eff}$ from equation~\ref{eq:coupling} and $\Omega_{\rm beam}$.

\begin{figure}
\centering
\includegraphics[width=\hsize]{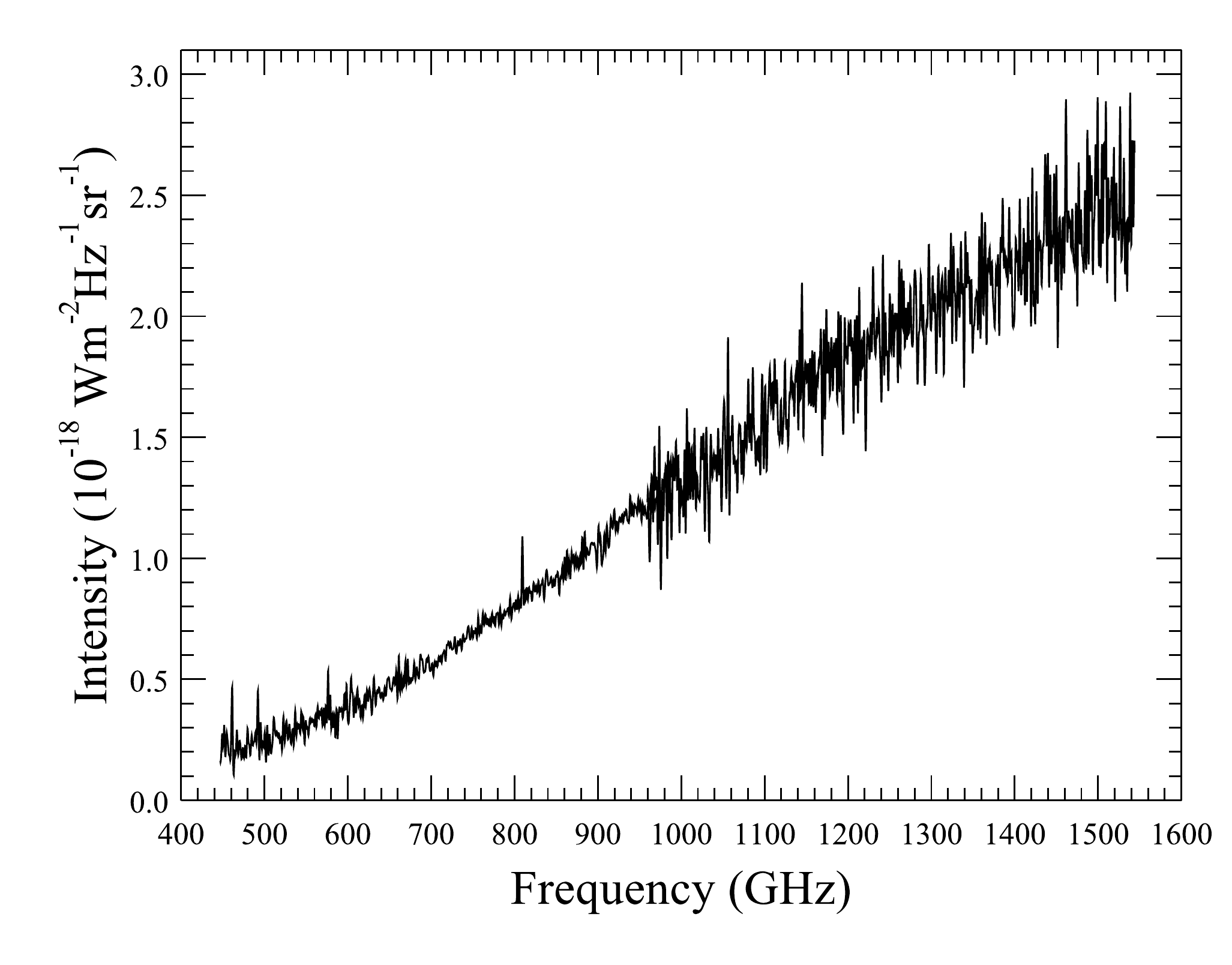}
\caption{Spectrum of Cas~A, an extended source, showing the agreement between the SLW and SSW calibration. The apparent increase in noise in the SSW portion of the spectrum is due to the smaller beam size compared to SLW.}
\label{fig:extendedspec}
\end{figure}

There is no independent measurement of the feedhorn coupling efficiency for SSW. However, there is excellent agreement observed in the overlap between SSW and SLW in spectra of extended sources (see Fig.~\ref{fig:extendedspec}) and this indicates that we can rely on the SLW estimate of $\Omega_{\rm eff}^\prime$. The derived form of the SSW feedhorn coupling efficiency required in order to make SSW consistent is approximately flat in frequency, with a value of 0.65 \citep{wu}. Fig.~\ref{fig:coupling} shows the resulting value of $\Omega_{\rm eff}^\prime$, which gives excellent agreement to the measured effective beam size, $\Omega_{\rm eff}$.

The error on the measured SLW coupling efficiency quoted by \citet{chattopadhyay-ieeetrans-2003} is $\pm$3\%. This error can be translated directly into the absolute calibration uncertainty assuming the simple diffraction model is correct. When it is combined with the uncertainty on the Uranus point source model from Section~\ref{sect:u_model}, the overall uncertainty on the absolute flux calibration for extended sources is 6\%. However, when the source is not fully extended, or if there is structure inside the beam, the uncertainties are dominated by the source-beam coupling \citep[see][]{wu} and are significantly greater than this.

\subsection{Systematic uncertainty in the instrument and telescope models}\label{sect:tel_model}

The systematic uncertainties on the instrument and telescope model terms combine as linear terms in the overall uncertainty estimation. In general the stochastic terms represent the uncertainty within a given spectral sample and between spectral samples, and will dominate the noise when considering faint spectral features.  However, the model terms dominate the error in the overall continuum level and spectral shape.

The telescope emission dominates the total signal of nearly all FTS observations, as the associated flux density within the beam is in the range $\sim$200--800\,Jy. After subtracting the telescope model, any residual that remains is extended in the beam and, therefore, on application of the point-source calibration, it manifests as a distortion of the overall spectral shape and a mismatch between the two bands. This discrepancy is due to the change in beam \'{e}tendue ($A\Omega$) with frequency. Residual emission from the instrument only makes a significant contribution to the associated uncertainty in SLW, and dominates below 600\,GHz. 

\begin{figure}
\centering
\includegraphics[width=\hsize]{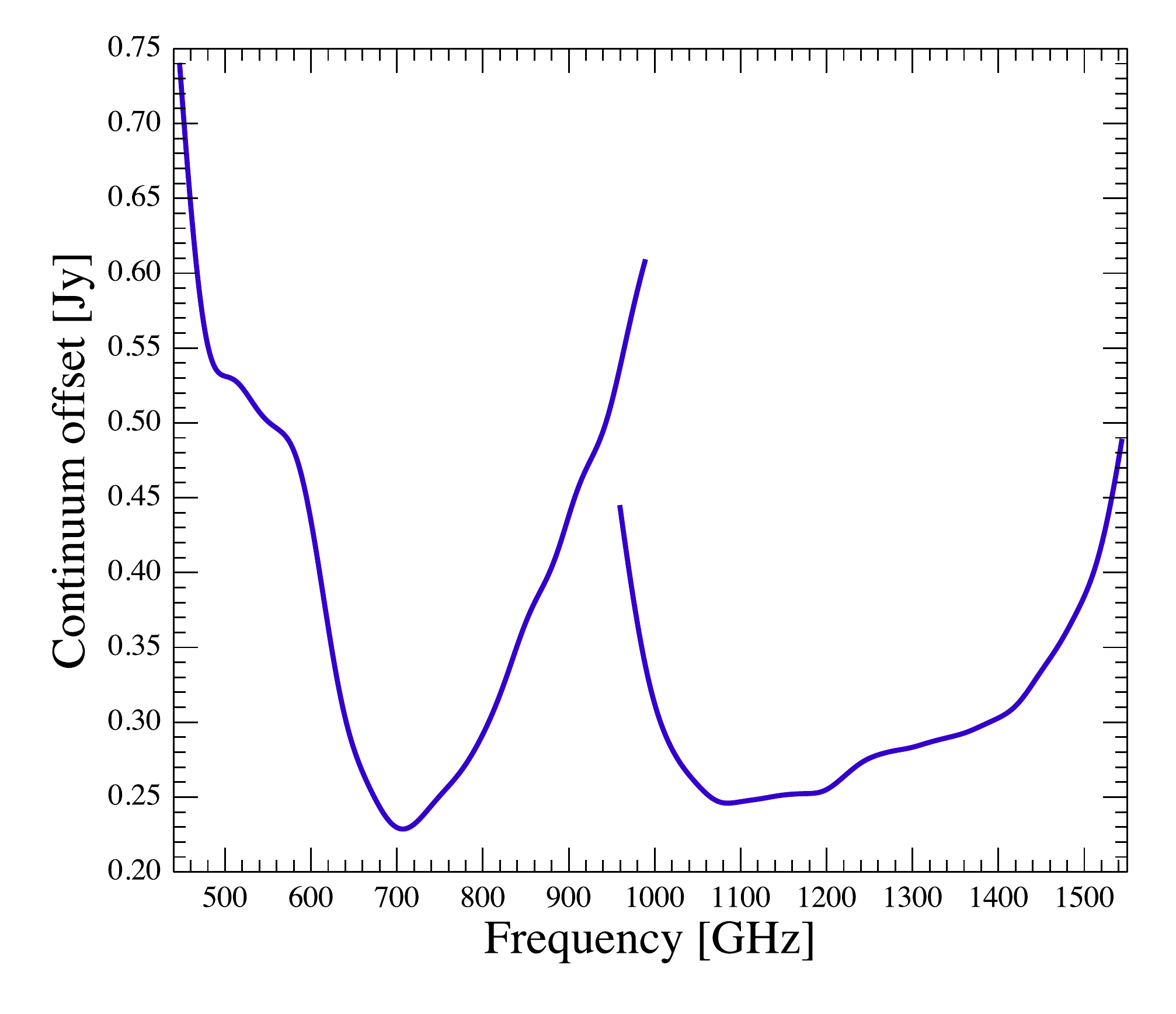}
\includegraphics[width=\hsize]{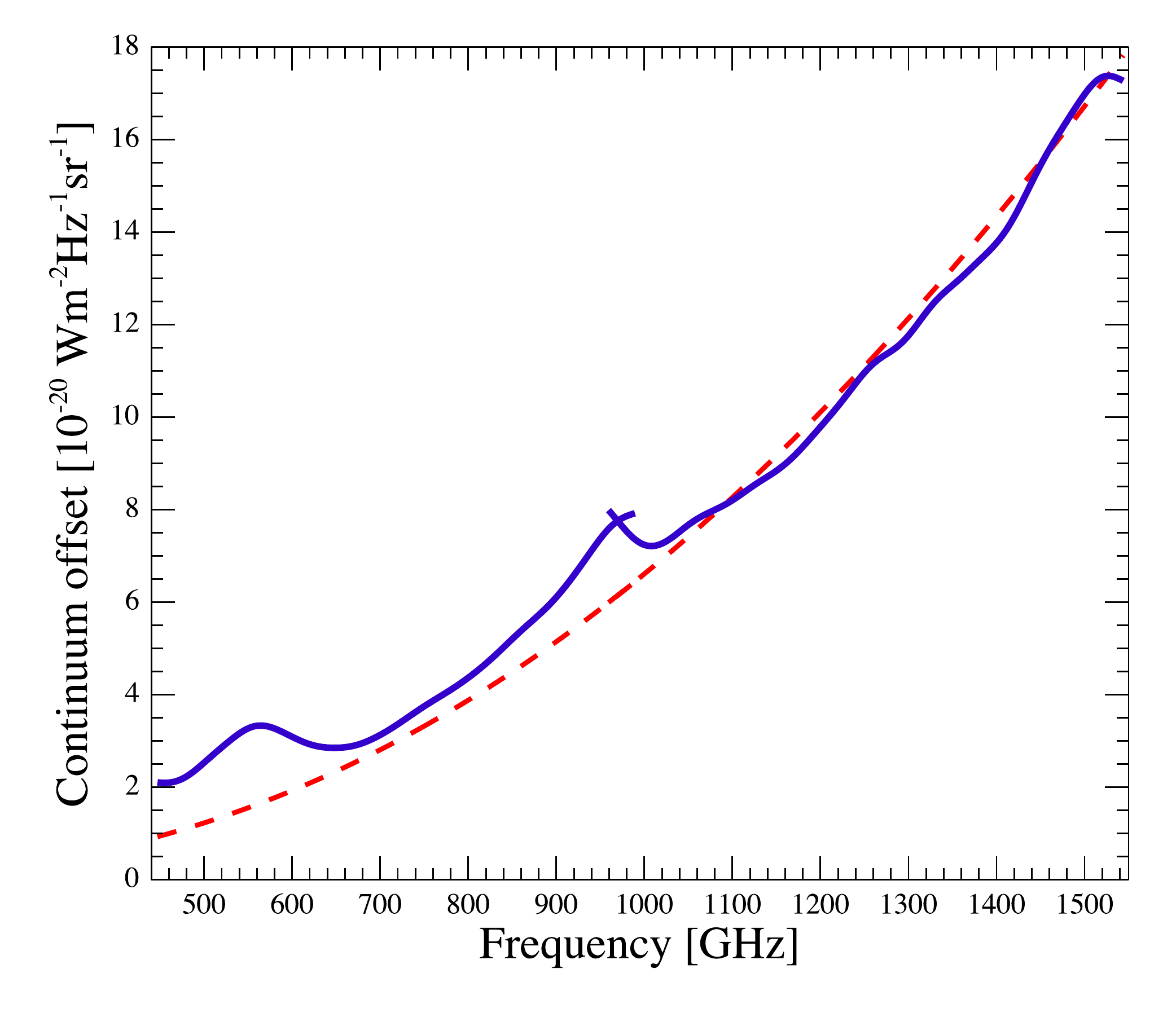}
\caption{The 1$\sigma$ additive continuum offset determined from the standard deviation of measured dark sky spectra over the whole mission, in point-source calibrated units (top) and extended-source calibrated units (bottom). The red dashed line shows that this can be understood as a 0.058\% uncertainty on the telescope model, with additional contribution below 600\,GHz due to uncertainty on the instrument model.}
\label{fig:offset}
\end{figure}

The expected uncertainty on the measured continuum level, due to this residual background, was assessed using the large scale spectral shape of dark sky observations. All of the long ($>$\,20 repetitions) dark sky observations were point-source calibrated, but not averaged, providing over 10,000 individual dark scans that cover the whole mission. Each scan was smoothed with a wide (21\,GHz) Gaussian kernel, and the spread obtained from the standard deviation across the entire set of smoothed scans. The top plot of Fig.~\ref{fig:offset} shows the resulting point-source calibrated 1$\sigma$ additive continuum offset for the centre detectors as a function of frequency. This additive offset is significantly less than 1\,Jy, with average values of 0.40\,Jy for SLWC3 and 0.29\,Jy for SSWD4. 

The bottom plot of Fig.~\ref{fig:offset} presents the offset in extended-source calibrated units. To illustrate what the offset represents as a fraction of the telescope emission, it is compared with 0.058\% of an average telescope model (red dashed line), constructed as the median of the models associated with the dark sky observations used to calculate the continuum offset. This fraction represents the estimated uncertainty on the telescope model across the frequency bands. Below 600\,GHz, a significant contribution to the continuum offset from the residual instrument emission can be seen in Fig.~\ref{fig:offset}.

\subsection{Spectral lines \label{sect:specLines}}

The accuracy with which integrated spectral line fluxes can be extracted is limited by both the overall calibration accuracy discussed in previous sections and the knowledge of the instrumental line shape used for fitting the line. If the instrument line shape was known perfectly, the calibration accuracy would be the same as for the continuum due to the nature of the FTS operation (both lines and continuum are measured together in the interferogram - e.g. see Fig.~\ref{interferogram}).

As described in Section~\ref{sect_freq}, the instrumental line shape can be fitted by a sinc function with a spectral resolution that varies in velocity between 230--800~kms$^{-1}$ across the band. This spectral resolution means that most Galactic sources have unresolved lines, and a sinc function fit is a good approximation. Extra-galactic targets with very broad lines may be better fitted with a convolution of a Gaussian and sinc function \citep[e.g.][]{rigopoulou}.

In order to test the limits of the spectral line fitting, routine calibration observations of NGC7027 were used (see also Section~\ref{sect:repeatability}). The $^{12}$CO lines in this source were observed with a signal-to-noise ratio of greater than 300. The line widths measured across the different spectral lines in all of the observations match the expected resolution, with a scatter consistent with the noise level of the data. 

\begin{figure}
\centering
\includegraphics[width=1.0\hsize]{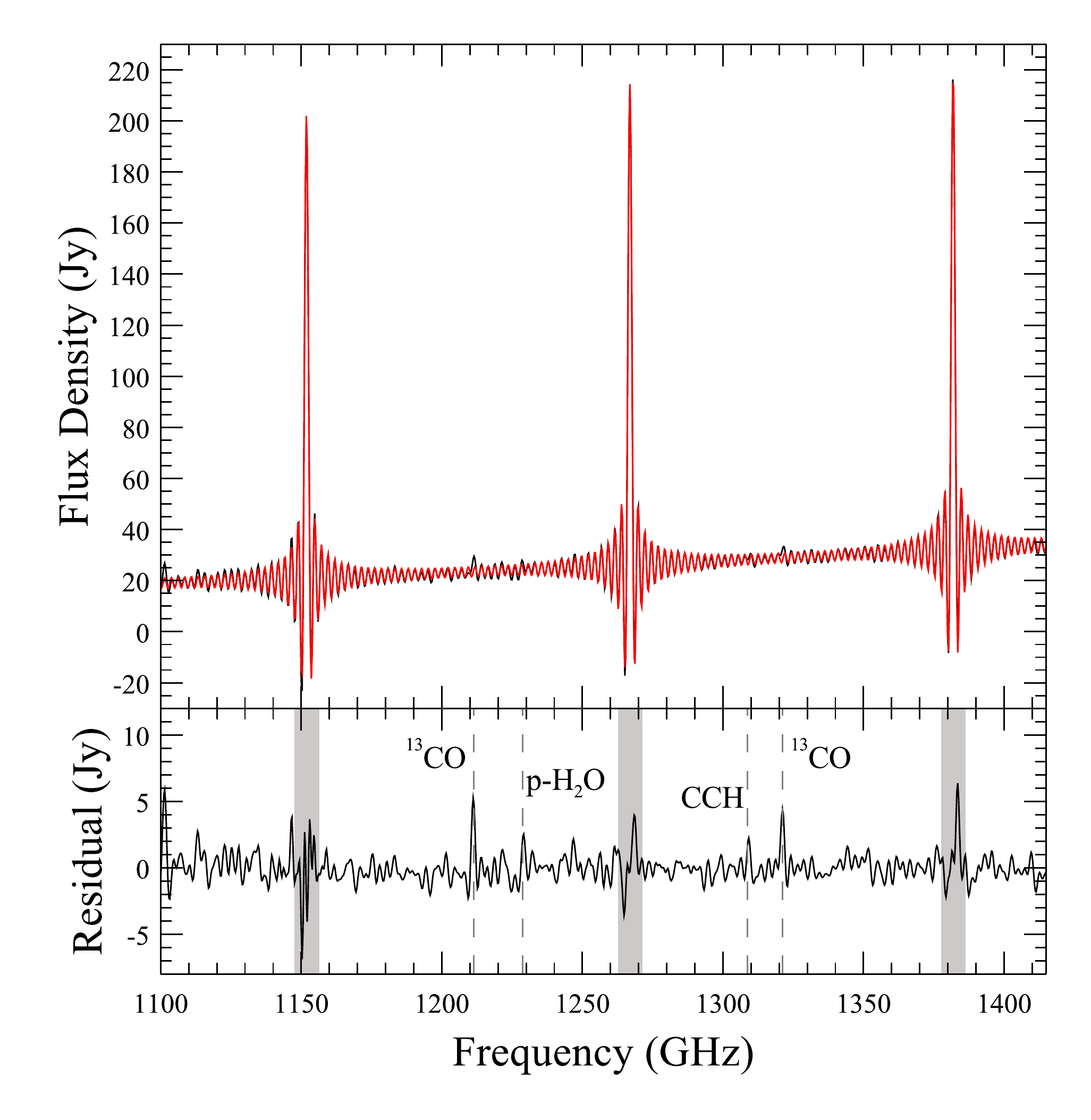}
\caption{Part of the fitted spectrum of NGC7027 showing the $^{12}$CO $J$=10--9, 11--10 and 12--11 lines. Once the main lines have been fitted, weaker features are visible in the residual, and several of these are marked \citep[see][]{wesson}. Some residual is also visible around the $^{12}$CO lines (grey bands).}
\label{fig:lines}
\end{figure}

Fig.~\ref{fig:lines} shows an example fit to the $^{12}$CO lines in one of the observations of NGC7027, with the lines and continuum fitted simultaneously and the line widths fixed to the instrument resolution (using a line fitting script that is available inside HIPE). The residual shows that the sinc line shape does not exactly match the shape of the $^{12}$CO lines in the data (at a level of a few percent). This mismatch is related to a systematic asymmetry in the first negative high frequency lobe of the instrument line shape\footnote{The use of an empirical, asymmetric instrument line shape is under investigation, and may be available in future versions of HIPE.} \citep[e.g. see][]{naylor10}, but could also be affected by any weaker spectral features that may be blended with the main lines. 

These systematic effects should be kept in mind when fitting spectral lines with high signal-to-noise ratio, and the residual should be investigated to determine the error on the fitted line flux. Once the main lines have been fitted, it may be possible to identify weaker features in the residual, although Fig.~\ref{fig:lines} shows that care must be taken to avoid misidentifications at the positions of the main lines themselves.

The standard FTS pipeline also produces spectra after an apodization function has been applied to the interferogram \citep[the adjusted Norton-Beer 1.5 function;][]{naylorTahic}, and these products appear in the Herschel Science Archive labelled as ``apodized''. This apodization function smooths the instrumental line shape to reduce the sinc wings, with a cost of degrading the spectral resolution by a factor of 1.5. The instrumental line shape can then be approximately fitted using a Gaussian (although the apodized line shape is not strictly Gaussian). The results recovered from a Gaussian fit of apodized data overestimate the line flux by up to 5\% compared to the sinc fit of the un-apodized data.

\subsection{Repeatability \label{sect:repeatability}}

Over the {\it Herschel} mission, the accuracy and repeatability of the SPIRE FTS calibration scheme was monitored with a programme of routine calibration observations. This programme included regular observations of the primary calibrator Uranus, the SPIRE dark sky region, secondary calibrators such as Neptune and the brighter asteroids, and a number of line sources such as AFGL2688 and NGC7027. 

Repeatability was assessed using line measurements of sources with strong spectral features -- AFGL2688, AFGL4106, CRL618 and NGC7027. These sources and their observations will be discussed in Hopwood et al. (in preparation). Excluding observations with known pointing offsets, the line flux and line velocity were measured for each set of observations, for each source, to assess the FTS repeatability over the whole mission. Random pointing errors (see Section \ref{sect:pointUn}) were not corrected and therefore, the repeatability values include the effect of pointing uncertainty. For the centre detectors the variation in measured line flux is found to be $<$6\% (Hopwood et al., in preparation).

\begin{figure}
\centering
\includegraphics[width=1.0\hsize]{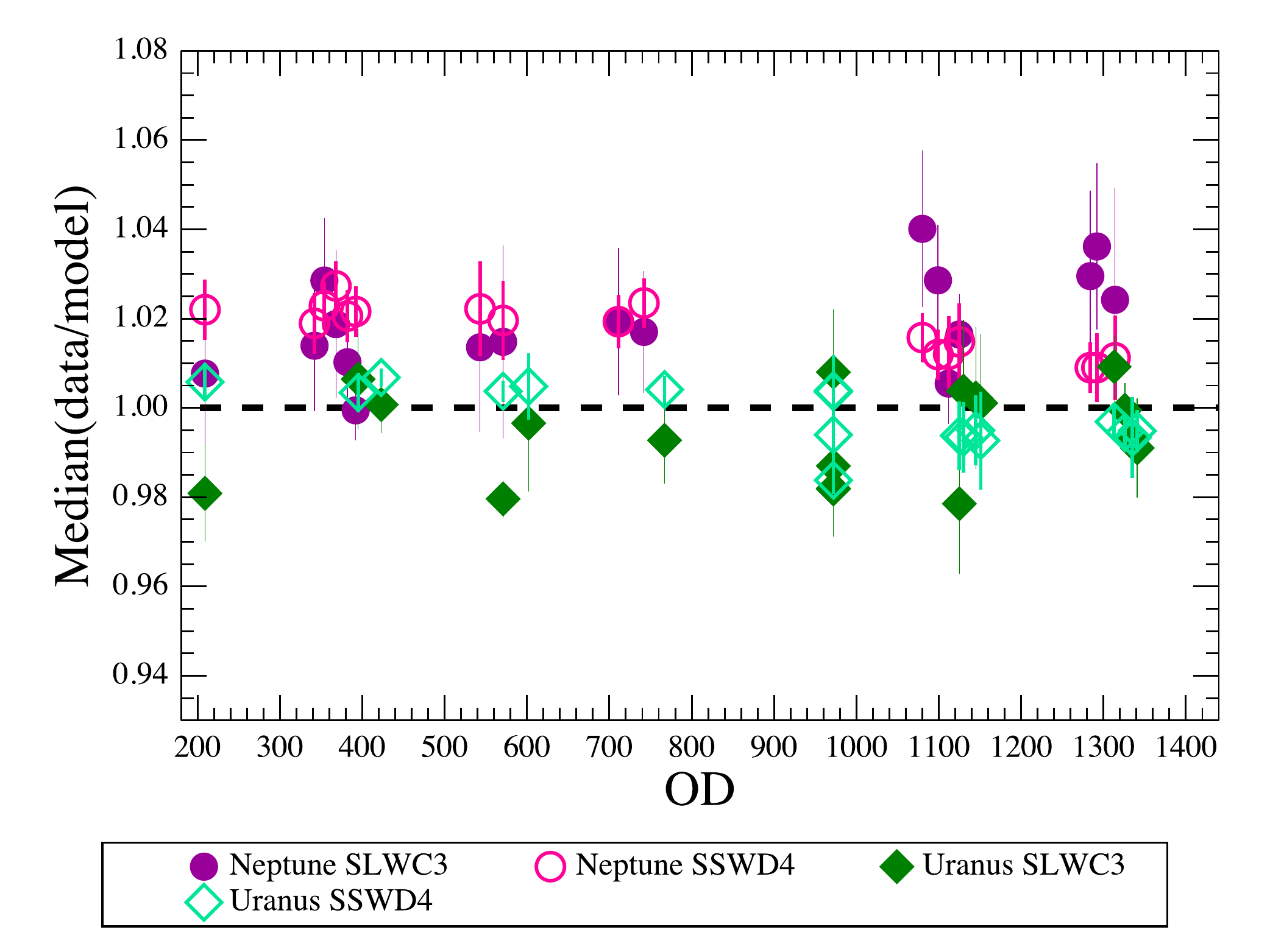}
\caption{Data-to-model ratios for observations of Neptune and Uranus. The median ratios are taken for pointing corrected data, excluding the noisy ends, i.e., over 500--950\,GHz for SLWC3 and 1100--1500\,GHz for SSWD4.}
\label{fig:planetRatios}
\end{figure}

The repeatability of the calibration can also be assessed by comparing the observed spectra of Uranus and Neptune with the model predictions. Fig.~\ref{fig:planetRatios} shows data-to-model ratios for Uranus and Neptune over the course of the mission. In this case, pointing effects can be taken into account \citep[see][]{valtchanov}, and each ratio is the median of pointing corrected data to the respective model. To avoid the noisy ends of each band the ratios were calculated over a truncated frequency range: 500--950~GHz for SLWC3 and 1100--1500~GHz for SSWD4. The Uranus ratios are scattered around 1.0 and show an overall agreement of better than 1\%. 

The repeatability of spectral mapping observations has been investigated using observations of the Orion Bar, and is described in detail by \citet{benielli}. They found that in fully sampled mapping mode, the deviation of repeated observations through the mission was $\sim$7\%, although the uncertainty rises to $\sim$9\% towards the edges of the $\sim$2$^{\prime}$ map.

\subsection{Consistency with SPIRE photometer}

An independent check of the FTS calibration can be made by comparing with the SPIRE photometer, as it is calibrated using Neptune. In order to minimise uncertainty due to the different beam sizes of the two instruments, the five most point-like FTS routine calibration sources, CRL618, NGC7027, CW Leo, R Dor, VYCMa (discussed in Hopwood et al., in preparation) were used to carry out a photometric comparison between the spectrometer and photometer. For each source there are one or more publicly available maps, which were obtained from the {\it Herschel} Science Archive, processed using the HIPE V10.3 photometer pipeline and equivalent calibration files (SPIRE\_CAL\_10\_1). The HIPE sourceExtractorTimeline task was used to obtain flux densities from the Level-1 data for each PSW (250~$\mu$m), PMW (350~$\mu$m) and PLW (500~$\mu$m) map. The resulting photometry for each photometric band was averaged over the multiple observations of each source, and the errors added in quadrature. There are multiple FTS observations for each of the seven sources, spanning the whole {\it Herschel} mission. The standard Level-2 spectrometer pipeline product was used to obtain the synthetic PSW, PMW and PLW photometry by  using the SPIRE photometer filters for each observation. The ratio of spectrometer to average photometer photometry was taken and the mean ratio and standard deviation over the repeated FTS observations for each photometer band are: PSW 1.02$\pm$0.06; PMW 0.96$\pm$0.07; PLW 1.04$\pm$0.06.

\subsection{Pointing uncertainties}
\label{sect:pointUn}

There is an additional source of photometric uncertainty for point sources,  originating from the absolute pointing error (APE) associated with all {\it Herschel} observations. The APE varied between 1--2\arcsec through the mission at the 1$\sigma$ 68\% confidence level, with significant improvement down to $0.9\arcsec$ for observations after OD\,1011 \citep{sanchez13}. The magnitude of the flux uncertainty, due to the APE, depends on frequency as it has a much larger effect on the 16.5\arcsec\ beam at the high-frequency end of SSW than the 37\arcsec\ beam at the low-frequency end of SLW. 

There is no systematic difference in the distribution of pointing offsets for calibration observations taken before and after the 1.7\arcsec\ change in the BSM offset \citep{valtchanov}, because the APE and the BSM offset are similar in magnitude, and the directions of both can be assumed to be arbitrary. Therefore, the 1$\sigma$ range in source position varies between the centre of the beam and an offset of 3.7\arcsec, assuming an APE of 2\arcsec. This range would introduce a drop of $\sim10$\% of the continuum level, which we consider as the 1$\sigma$ limit on the uncertainty introduced by the pointing APE for observations before changing the BSM mirror position. For observations after the change, the APE would introduce a 1$\sigma$ reduction in flux density of 4\%. \citet{valtchanov} measured the actual pointing offset for 29 observations of Uranus on the centre detectors, and found that the worst case offset was 4.2\arcsec.

There is another pointing uncertainty, the relative pointing error (RPE), which is a measure of the stability of the telescope once commanded to a given sky position. The RPE is estimated to be better than the required pre-launch value of 0.3\arcsec\  \citep{sanchez13} and this has no noticeable effects on the flux levels.

\section{Summary}

The overall absolute calibration uncertainty for the FTS nominal mode is summarised in this section. We break the uncertainties into three categories: point sources observed in the sparse mode, extended sources observed in the sparse mode, and extended sources observed in mapping mode. The calibration described in this paper has been implemented in the pipeline corresponding to HIPE Version 11.

To summarise for point sources observed on the centre detectors (SSWD4 and SLWC3), the measured repeatability is 6\%, with the following contributions: (i) absolute systematic uncertainty in the models from comparison of Uranus and Neptune - determined to be $\pm$3\%; (ii) the statistical repeatability determined from observations of Uranus and Neptune, with pointing corrected - estimated at $\pm$1\% (excluding the edges of the bands); (iii) the uncertainties in the instrument and telescope model, which lead to an additive continuum offset error of 0.4\,Jy for SLW and 0.3\,Jy for SSW and (iv) the effect of the {\it Herschel} APE.

Note that the pointing uncertainty results in a reduction in flux and is, therefore, a one-sided statistical uncertainty on the calibrated spectrum. A large pointing offset also results in a significant distortion of the SSW spectrum of a point source and a mismatch between the SLW and SSW spectra \citep[e.g. see][]{valtchanov}. Providing one is convinced that the source in question has no spatial extent, the SLW portion of the calibrated spectrum can be used to correct any apparent gain difference between the SLW and SSW spectra.

For sparse observations of significantly extended sources, the absolute uncertainty in intensity for a reasonably bright, fully extended object, observed in the central detectors is 7\%, with the following contributions: (i) the uncertainty in comparing the calibration on Uranus (a point source) to the telescope is estimated at 3\%; (ii) the uncertainty on the Uranus model itself of 3\%; (iii) the systematic reproducibility of the telescope model of 0.06\%; (iv) the statistical repeatability estimated at $\pm$1\% and (v) an additive continuum offset of $3.4\times10^{-20}$\,\wmhzsr~for SLW and $1.1\times10^{-19}$\,\wmhzsr~for SSW.

In practice, truly extended sources tend to be faint and the uncertainty is therefore dominated by the additive offsets. When the source extent is larger than the main beam size, but not fully extended, or if there is structure inside the beam, then the uncertainties are dominated by the source-beam coupling \citep[see][]{wu} and are significantly greater.

In mapping mode, the variations between detectors becomes important and the overall repeatability has been measured as $\pm$7\% \citep[see][for a full discussion of mapping mode observations]{benielli}. The off-axis detectors are less well calibrated, especially outside the unvignetted part of the field.

The level of absolute flux accuracy and repeatability obtained with the SPIRE FTS compares favourably with the SPIRE photometer \citep{bendo}.

The excellent level of calibration accuracy achieved is due to the linear transform properties of the FTS, which guarantee simultaneous calibration of the entire spectrum. Thus spectral features covering a wide range in frequency can be analysed together.  This is not possible with monochromating devices where only a narrow frequency range is observed, as that can lead to calibration uncertainties caused by spectral features that are broader than, or outside of, the observed band (for instance standing waves or broad spectral features in the instrument response function). The penalty is in instantaneous sensitivity, due to the increased photon noise in an FTS. However, this noise is compensated for by the higher level of calibration  fidelity that can be achieved, as well as the more widely appreciated advantage in observing speed for multi-line spectral observations.

\section*{Acknowledgments}
We thank the referee for their thoughtful comments which helped improve the paper.

SPIRE has been developed by a consortium of institutes led by Cardiff Univ. (UK) and including: Univ. Lethbridge (Canada); NAOC (China); CEA, LAM (France); IFSI, Univ. Padua (Italy); IAC (Spain); Stockholm Observatory (Sweden); Imperial College London, RAL, UCL-MSSL, UKATC, Univ. Sussex (UK); and Caltech, JPL, NHSC, Univ. Colorado (USA). This development has been supported by national funding agencies: CSA (Canada); NAOC (China); CEA, CNES, CNRS (France); ASI (Italy); MCINN (Spain); SNSB (Sweden); STFC, UKSA (UK); and NASA (USA). 

N.M. is funded by an ASI fellowship under contract number I/005/11/0.

\bibliographystyle{aa}
\bibliography{references}

\begin{thebibliography}{49}
\expandafter\ifx\csname natexlab\endcsname\relax\def\natexlab#1{#1}\fi

\bibitem[{Ade {et~al.}(1999)Ade, Hamilton, \& Naylor}]{ade1999}
Ade, P., Hamilton, P., \& Naylor, D. 1999, in {Fourier Transform Spectroscopy:
  New Methods and Applications}, Optical Society of America, FWE3

\bibitem[{{Bendo} {et~al.}(2013){Bendo}, {Griffin}, {Bock}, {Conversi},
  {Dowell}, {Lim}, {Lu}, {North}, {Papageorgiou}, {Pearson}, {Pohlen},
  {Polehampton}, {Schulz}, {Shupe}, {Sibthorpe}, {Spencer}, {Swinyard},
  {Valtchanov}, \& {Xu}}]{bendo}
{Bendo}, G.~J., {Griffin}, M.~J., {Bock}, J.~J., {et~al.} 2013, MNRAS, 433,
  3062

\bibitem[{Benielli {et~al.}(2014)Benielli, Polehampton, Hopwood,
  {Gri\~{n}\'{o}n Mar\'{i}n}, Fulton, Imhof, Lim, Lu, Marchili, Naylor,
  Swinyard, \& Valtchanov}]{benielli}
Benielli, D., Polehampton, E., Hopwood, R., {et~al.} 2014, Exp. Ast.

\bibitem[{{Buckle} {et~al.}(2012){Buckle}, {Davis}, {Francesco}, {Graves},
  {Nutter}, {Richer}, {Roberts}, {Ward-Thompson}, {White}, {Brunt}, {Butner},
  {Cavanagh}, {Chrysostomou}, {Curtis}, {Duarte-Cabral}, {Etxaluze}, {Fich},
  {Friberg}, {Friesen}, {Fuller}, {Greaves}, {Hatchell}, {Hogerheijde},
  {Johnstone}, {Matthews}, {Matthews}, {Rawlings}, {Sadavoy}, {Simpson},
  {Tothill}, {Tsamis}, {Viti}, {Wouterloot}, \& {Yates}}]{buckle}
{Buckle}, J.~V., {Davis}, C.~J., {Francesco}, J.~D., {et~al.} 2012, MNRAS, 422,
  521

\bibitem[{Chattopadhyay {et~al.}(2003)Chattopadhyay, Glenn, Bock, Rownd,
  Caldwell, \& Griffin}]{chattopadhyay-ieeetrans-2003}
Chattopadhyay, G., Glenn, J., Bock, J., {et~al.} 2003, Microwave Theory and
  Techniques, IEEE Transactions on, 51, 2139

\bibitem[{{Conrath} {et~al.}(1987){Conrath}, {Hanel}, {Gautier}, {Marten}, \&
  {Lindal}}]{conrath}
{Conrath}, B., {Hanel}, R., {Gautier}, D., {Marten}, A., \& {Lindal}, G. 1987,
  J. Geophys. Res., 92, 15003

\bibitem[{{Dohlen} {et~al.}(2000){Dohlen}, {Origne}, {Pouliquen}, \&
  {Swinyard}}]{dohlen2000}
{Dohlen}, K., {Origne}, A., {Pouliquen}, D., \& {Swinyard}, B.~M. 2000, in
  Society of Photo-Optical Instrumentation Engineers (SPIE) Conference Series,
  Vol. 4013, Society of Photo-Optical Instrumentation Engineers (SPIE)
  Conference Series, ed. J.~B. {Breckinridge} \& P.~{Jakobsen}, 119--128

\bibitem[{{Feuchtgruber} {et~al.}(2013){Feuchtgruber}, {Lellouch}, {Orton}, {de
  Graauw}, {Vandenbussche}, {Swinyard}, {Moreno}, {Jarchow}, {Billebaud},
  {Cavali{\'e}}, {Sidher}, \& {Hartogh}}]{feuchtgruber}
{Feuchtgruber}, H., {Lellouch}, E., {Orton}, G., {et~al.} 2013, A\&A, 551, A126

\bibitem[{{Fischer} {et~al.}(2004){Fischer}, {Klaassen}, {Hovenier}, {Jakob},
  {Poglitsch}, \& {Sternberg}}]{fischer04}
{Fischer}, J., {Klaassen}, T., {Hovenier}, N., {et~al.} 2004, Appl. Opt., 43,
  3765

\bibitem[{Fulton {et~al.}(2014)Fulton, Hopwood, Baluteau, Benielli, Imhof, Lim,
  Lu, Marchili, Naylor, Polehampton, Swinyard, \& Valtchanov}]{fultonExpAst}
Fulton, T., Hopwood, R., Baluteau, J.-P., {et~al.} 2014, Exp. Ast.

\bibitem[{{Giorgini} {et~al.}(1996){Giorgini}, {Yeomans}, {Chamberlin},
  {Chodas}, {Jacobson}, {Keesey}, {Lieske}, {Ostro}, {Standish}, \&
  {Wimberly}}]{giorgini}
{Giorgini}, J.~D., {Yeomans}, D.~K., {Chamberlin}, A.~B., {et~al.} 1996, in
  Bulletin of the American Astronomical Society, Vol.~28, AAS/Division for
  Planetary Sciences Meeting Abstracts \#28, 1158

\bibitem[{Griffin(2007)}]{sensitivityNote}
Griffin, M.~J. 2007, SPIRE Sensitivity Models, research report
  SPIRE-QMW-NOT-000642, SPIRE Consortium, Harwell

\bibitem[{{Griffin} {et~al.}(2010){Griffin}, {Abergel}, {Abreu}, {Ade},
  {Andr{\'e}}, {Augueres}, {Babbedge}, {Bae}, {Baillie}, {Baluteau}, {Barlow},
  {Bendo}, {Benielli}, {Bock}, {Bonhomme}, {Brisbin},
  {Brock2004ApOpt..43.3765Fley-Blatt}, {Caldwell}, {Cara}, {Castro-Rodriguez},
  {Cerulli}, {Chanial}, {Chen}, {Clark}, {Clements}, {Clerc}, {Coker},
  {Communal}, {Conversi}, {Cox}, {Crumb}, {Cunningham}, {Daly}, {Davis}, {de
  Antoni}, {Delderfield}, {Devin}, {di Giorgio}, {Didschuns}, {Dohlen},
  {Donati}, {Dowell}, {Dowell}, {Duband}, {Dumaye}, {Emery}, {Ferlet},
  {Ferrand}, {Fontignie}, {Fox}, {Franceschini}, {Frerking}, {Fulton},
  {Garcia}, {Gastaud}, {Gear}, {Glenn}, {Goizel}, {Griffin}, {Grundy}, {Guest},
  {Guillemet}, {Hargrave}, {Harwit}, {Hastings}, {Hatziminaoglou}, {Herman},
  {Hinde}, {Hristov}, {Huang}, {Imhof}, {Isaak}, {Israelsson}, {Ivison},
  {Jennings}, {Kiernan}, {King}, {Lange}, {Latter}, {Laurent}, {Laurent},
  {Leeks}, {Lellouch}, {Levenson}, {Li}, {Li}, {Lilienthal}, {Lim}, {Liu},
  {Lu}, {Madden}, {Mainetti}, {Marliani}, {McKay}, {Mercier}, {Molinari},
  {Morris}, {Moseley}, {Mulder}, {Mur}, {Naylor}, {Nguyen}, {O'Halloran},
  {Oliver}, {Olofsson}, {Olofsson}, {Orfei}, {Page}, {Pain}, {Panuzzo},
  {Papageorgiou}, {Parks}, {Parr-Burman}, {Pearce}, {Pearson},
  {P{\'e}rez-Fournon}, {Pinsard}, {Pisano}, {Podosek}, {Pohlen}, {Polehampton},
  {Pouliquen}, {Rigopoulou}, {Rizzo}, {Roseboom}, {Roussel}, {Rowan-Robinson},
  {Rownd}, {Saraceno}, {Sauvage}, {Savage}, {Savini}, {Sawyer}, {Scharmberg},
  {Schmitt}, {Schneider}, {Schulz}, {Schwartz}, {Shafer}, {Shupe}, {Sibthorpe},
  {Sidher}, {Smith}, {Smith}, {Smith}, {Spencer}, {Stobie}, {Sudiwala},
  {Sukhatme}, {Surace}, {Stevens}, {Swinyard}, {Trichas}, {Tourette}, {Triou},
  {Tseng}, {Tucker}, {Turner}, {Vaccari}, {Valtchanov}, {Vigroux}, {Virique},
  {Voellmer}, {Walker}, {Ward}, {Waskett}, {Weilert}, {Wesson}, {White},
  {Whitehouse}, {Wilson}, {Winter}, {Woodcraft}, {Wright}, {Xu}, {Zavagno},
  {Zemcov}, {Zhang}, \& {Zonca}}]{griffin2010}
{Griffin}, M.~J., {Abergel}, A., {Abreu}, A., {et~al.} 2010, \aap, 518, L3

\bibitem[{{Griffin} \& {Orton}(1993)}]{griffinOrton}
{Griffin}, M.~J. \& {Orton}, G.~S. 1993, Icarus, 105, 537

\bibitem[{{Hargrave} {et~al.}(2006){Hargrave}, {Waskett}, {Lim}, \&
  {Swinyard}}]{hargrave2006}
{Hargrave}, P., {Waskett}, T., {Lim}, T., \& {Swinyard}, B. 2006, in Society of
  Photo-Optical Instrumentation Engineers (SPIE) Conference Series, Vol. 6275,
  Society of Photo-Optical Instrumentation Engineers (SPIE) Conference Series

\bibitem[{{Herpin} {et~al.}(2002){Herpin}, {Goicoechea}, {Pardo}, \&
  {Cernicharo}}]{herpin}
{Herpin}, F., {Goicoechea}, J.~R., {Pardo}, J.~R., \& {Cernicharo}, J. 2002,
  ApJ, 577, 961

\bibitem[{{Hopwood} {et~al.}(2013){Hopwood}, {Fulton}, {Polehampton},
  {Valtchanov}, {Benielli}, {Imhof}, {Lim}, {Lu}, {Marchili}, {Pearson}, \&
  {Swinyard}}]{hopwood}
{Hopwood}, R., {Fulton}, T., {Polehampton}, E.~T., {et~al.} 2013, Exp. Ast.

\bibitem[{{Josselin} {et~al.}(1998){Josselin}, {Loup}, {Omont}, {Barnbaum},
  {Nyman}, \& {Sevre}}]{josselin}
{Josselin}, E., {Loup}, C., {Omont}, A., {et~al.} 1998, A\&AS, 129, 45

\bibitem[{{Lu} {et~al.}(2013){Lu}, {Polehampton}, {Swinyard}, {Benielli},
  {Fulton}, {Hopwood}, {Imhof}, {Lim}, {Marchili}, {Naylor}, {Schulz},
  {Sidher}, \& {Valtchanov}}]{lu}
{Lu}, N., {Polehampton}, E.~T., {Swinyard}, B.~M., {et~al.} 2013, Exp. Ast.

\bibitem[{Mach(1892)}]{mach}
Mach, L. 1892, Zeitschrift f\"{u}r Instrumentenkunde, 12, 89

\bibitem[{Makiwa {et~al.}(2013)Makiwa, Naylor, Ferlet, Salji, Swinyard,
  Polehampton, \& {van der Wiel}}]{makiwa2013}
Makiwa, G., Naylor, D.~A., Ferlet, M., {et~al.} 2013, Appl. Opt., 52, 3864

\bibitem[{Mather(1982)}]{mather}
Mather, J.~C. 1982, Applied Optics, 21, 1125

\bibitem[{{Mellon} {et~al.}(2000){Mellon}, {Jakosky}, {Kieffer}, \&
  {Christensen}}]{mellon}
{Mellon}, M.~T., {Jakosky}, B.~M., {Kieffer}, H.~H., \& {Christensen}, P.~R.
  2000, Icarus, 148, 437

\bibitem[{Moreno(1998)}]{moreno}
Moreno, R. 1998, PhD thesis, Universit\'{e} de Paris

\bibitem[{{Murphy} \& Padman(1991)}]{murphy-irphys-1991}
{Murphy}, J.~A. \& Padman, R. 1991, Infrared Physics, 31, 291

\bibitem[{{Naylor} {et~al.}(2010){Naylor}, {Baluteau}, {Barlow}, {Benielli},
  {Ferlet}, {Fulton}, {Griffin}, {Grundy}, {Imhof}, {Jones}, {King}, {Leeks},
  {Lim}, {Lu}, {Makiwa}, {Polehampton}, {Savini}, {Sidher}, {Spencer},
  {Surace}, {Swinyard}, \& {Wesson}}]{naylor10}
{Naylor}, D.~A., {Baluteau}, J.-P., {Barlow}, M.~J., {et~al.} 2010, in Society
  of Photo-Optical Instrumentation Engineers (SPIE) Conference Series, Vol.
  7731, Society of Photo-Optical Instrumentation Engineers (SPIE) Conference
  Series

\bibitem[{Naylor \& Tahic(2007)}]{naylorTahic}
Naylor, D.~A. \& Tahic, M.~K. 2007, J. Opt. Soc. Am. A., 24, 3644

\bibitem[{{Nguyen} {et~al.}(2004){Nguyen}, {Bock}, {Ringold}, {Battle},
  {Elliott}, {Turner}, {Weilert}, {Hristov}, {Schulz}, {Ganga}, {Zhang},
  {Beeman}, {Ade}, \& {Hargrave}}]{hien2004}
{Nguyen}, H.~T., {Bock}, J.~J., {Ringold}, P., {et~al.} 2004, in Society of
  Photo-Optical Instrumentation Engineers (SPIE) Conference Series, Vol. 5498,
  Society of Photo-Optical Instrumentation Engineers (SPIE) Conference Series,
  ed. C.~M. {Bradford}, P.~A.~R. {Ade}, J.~E. {Aguirre}, J.~J. {Bock},
  M.~{Dragovan}, L.~{Duband}, L.~{Earle}, J.~{Glenn}, H.~{Matsuhara}, B.~J.
  {Naylor}, H.~T. {Nguyen}, M.~{Yun}, \& J.~{Zmuidzinas}, 196--207

\bibitem[{Orton {et~al.}(2014)Orton, Fletcher, Moses, Mainzer, Hines, Hammel,
  Martin-Torres, Burgdorf, Merlet, \& Line}]{orton}
Orton, G.~S., Fletcher, L.~N., Moses, J.~I., {et~al.} 2014, Icarus, in review

\bibitem[{{Ott}(2010)}]{ott}
{Ott}, S. 2010, in Astronomical Society of the Pacific Conference Series, Vol.
  434, Astronomical Data Analysis Software and Systems XIX, ed. Y.~{Mizumoto},
  K.-I. {Morita}, \& M.~{Ohishi}, 139

\bibitem[{{Pilbratt} {et~al.}(2010){Pilbratt}, {Riedinger}, {Passvogel},
  {Crone}, {Doyle}, {Gageur}, {Heras}, {Jewell}, {Metcalfe}, {Ott}, \&
  {Schmidt}}]{pilbratt10}
{Pilbratt}, G.~L., {Riedinger}, J.~R., {Passvogel}, T., {et~al.} 2010, \aap,
  518, L1

\bibitem[{{Pisano} {et~al.}(2005){Pisano}, {Hargrave}, {Griffin}, {Collins},
  {Beeman}, \& {Hermoso}}]{pisano}
{Pisano}, G., {Hargrave}, P., {Griffin}, M., {et~al.} 2005, Appl. Opt., 44,
  3208

\bibitem[{{Putzig} \& {Mellon}(2007)}]{putzig}
{Putzig}, N.~E. \& {Mellon}, M.~T. 2007, Icarus, 191, 68

\bibitem[{{Rigopoulou} {et~al.}(2014){Rigopoulou}, {Hopwood}, {Magdis},
  {Thatte}, {Swinyard}, {Farrah}, {Huang}, {Alonso-Herrero}, {Bock},
  {Clements}, {Cooray}, {Griffin}, {Oliver}, {Pearson}, {Riechers}, {Scott},
  {Smith}, {Vaccari}, {Valtchanov}, \& {Wang}}]{rigopoulou}
{Rigopoulou}, D., {Hopwood}, R., {Magdis}, G.~E., {et~al.} 2014, ApJ, 781, L15

\bibitem[{{Rudy} {et~al.}(1987){Rudy}, {Muhleman}, {Berge}, {Jakosky}, \&
  {Christensen}}]{rudy}
{Rudy}, D.~J., {Muhleman}, D.~O., {Berge}, G.~L., {Jakosky}, B.~M., \&
  {Christensen}, P.~R. 1987, Icarus, 71, 159

\bibitem[{{S\'{a}nchez-Portal} {et~al.}(2014, submitted){S\'{a}nchez-Portal},
  Marston, Altieri, Aussel, Feuchtgruber, Klaas, Linz, Lutz, Mer\'{\i}n,
  M\"uller, Nielbock, Oort, Pilbratt, Stephenson, Tuttlebee, \& {The Herschel
  Pointing Working Group}}]{sanchez13}
{S\'{a}nchez-Portal}, M., Marston, A., Altieri, B., {et~al.} 2014, submitted,
  Exp. Ast.

\bibitem[{{Serabyn} \& {Weisstein}(1996)}]{serabyn}
{Serabyn}, E. \& {Weisstein}, E.~W. 1996, Applied Optics, 35, 2752

\bibitem[{Spencer {et~al.}(2010)Spencer, Naylor, \& Swinyard}]{spencer}
Spencer, L.~D., Naylor, D.~A., \& Swinyard, B.~M. 2010, Meas. Sci. Technol.,
  21, 065601

\bibitem[{{SPIRE Observer's Manual}(2014)}]{observersmanual}
{SPIRE Observer's Manual}. 2014, {HERSCHEL-HSC-DOC-0798}, accessed from
  http://herschel.esac.esa.int/Documentation.shtml

\bibitem[{Sudiwala {et~al.}(2002)Sudiwala, Griffin, \& Woodcraft}]{sudiwala}
Sudiwala, R.~V., Griffin, M.~J., \& Woodcraft, A.~L. 2002, Int. J. Infrared.
  Mm. Waves, 23, 575

\bibitem[{{Swinyard} {et~al.}(2010){Swinyard}, {Ade}, {Baluteau}, {Aussel},
  {Barlow}, {Bendo}, {Benielli}, {Bock}, {Brisbin}, {Conley}, {Conversi},
  {Dowell}, {Dowell}, {Ferlet}, {Fulton}, {Glenn}, {Glauser}, {Griffin},
  {Griffin}, {Guest}, {Imhof}, {Isaak}, {Jones}, {King}, {Leeks}, {Levenson},
  {Lim}, {Lu}, {Makiwa}, {Naylor}, {Nguyen}, {Oliver}, {Panuzzo},
  {Papageorgiou}, {Pearson}, {Pohlen}, {Polehampton}, {Pouliquen},
  {Rigopoulou}, {Ronayette}, {Roussel}, {Rykala}, {Savini}, {Schulz},
  {Schwartz}, {Shupe}, {Sibthorpe}, {Sidher}, {Smith}, {Spencer}, {Trichas},
  {Triou}, {Valtchanov}, {Wesson}, {Woodcraft}, {Xu}, {Zemcov}, \&
  {Zhang}}]{swinyard2010}
{Swinyard}, B.~M., {Ade}, P., {Baluteau}, J.-P., {et~al.} 2010, \aap, 518, L4

\bibitem[{{Swinyard} {et~al.}(2003){Swinyard}, {Dohlen}, {Ferand}, {Baluteau},
  {Pouliquen}, {Dargent}, {Michel}, {Martignac}, {Ade}, {Hargrave}, {Griffin},
  {Jennings}, \& {Caldwell}}]{swinyardSpie}
{Swinyard}, B.~M., {Dohlen}, K., {Ferand}, D., {et~al.} 2003, in Society of
  Photo-Optical Instrumentation Engineers (SPIE) Conference Series, Vol. 4850,
  Society of Photo-Optical Instrumentation Engineers (SPIE) Conference Series,
  ed. J.~C. {Mather}, 698--709

\bibitem[{{Teyssier} {et~al.}(2006){Teyssier}, {Hernandez}, {Bujarrabal},
  {Yoshida}, \& {Phillips}}]{teyssier}
{Teyssier}, D., {Hernandez}, R., {Bujarrabal}, V., {Yoshida}, H., \&
  {Phillips}, T.~G. 2006, A\&A, 450, 167

\bibitem[{Treffers(1977)}]{treffers}
Treffers, R. 1977, Applied Optics, 16, 3103

\bibitem[{{Turner} {et~al.}(2001){Turner}, {Bock}, {Beeman}, {Glenn},
  {Hargrave}, {Hristov}, {Nguyen}, {Rahman}, {Sethuraman}, \&
  {Woodcraft}}]{turner2001}
{Turner}, A.~D., {Bock}, J.~J., {Beeman}, J.~W., {et~al.} 2001, Appl. Opt., 40,
  4921

\bibitem[{{Valtchanov} {et~al.}(2013){Valtchanov}, {Hopwood}, {Polehampton},
  {Benielli}, {Fulton}, {Imhof}, {Konopczy{\'n}ski}, {Lim}, {Lu}, {Marchili},
  {Naylor}, \& {Swinyard}}]{valtchanov}
{Valtchanov}, I., {Hopwood}, R., {Polehampton}, E., {et~al.} 2013, Exp. Ast.

\bibitem[{{Wesson} {et~al.}(2010){Wesson}, {Cernicharo}, {Barlow}, {Matsuura},
  {Decin}, {Groenewegen}, {Polehampton}, {Agundez}, {Cohen}, {Daniel}, {Exter},
  {Gear}, {Gomez}, {Hargrave}, {Imhof}, {Ivison}, {Leeks}, {Lim}, {Olofsson},
  {Savini}, {Sibthorpe}, {Swinyard}, {Ueta}, {Witherick}, \& {Yates}}]{wesson}
{Wesson}, R., {Cernicharo}, J., {Barlow}, M.~J., {et~al.} 2010, 518, L144

\bibitem[{{Wu} {et~al.}(2013){Wu}, {Polehampton}, {Etxaluze}, {Makiwa},
  {Naylor}, {Salji}, {Swinyard}, {Ferlet}, {van der Wiel}, {Smith}, {Fulton},
  {Griffin}, {Baluteau}, {Benielli}, {Glenn}, {Hopwood}, {Imhof}, {Lim}, {Lu},
  {Panuzzo}, {Pearson}, {Sidher}, \& {Valtchanov}}]{wu}
{Wu}, R., {Polehampton}, E.~T., {Etxaluze}, M., {et~al.} 2013, A\&A, 556, 116

\bibitem[{Zehnder(1891)}]{zehnder}
Zehnder, L. 1891, Zeitschrift f\"{u}r Instrumentenkunde, 11, 275

\end{thebibliography}

\section*{Appendix: Details of the Uranus observations used for point-source calibration \label{appendix:uranus}}

The observations of Uranus used for deriving the point-source calibration are detailed in Table~\ref{tab_uranus}. Different observations were used for for high and low resolution observations, the two epochs before and after the BSM position was changed and different detectors. Where necessary, the listed dark sky observation was subtracted from the Uranus data to reduce systematic noise. The pointing offsets were derived for each observation by \citet{valtchanov}. For the off-axis detectors, there is no separation into the two BSM epochs because relevant observations were not made in both time periods.

\begin{table*}
\caption{Details of the Uranus observations used for point-source calibration. See \citet{valtchanov} for details on the pointing offsets.}
\medskip
\begin{center}
\begin{tabular}{llllll}
\hline\hline
Detector & Date & Repetitions & Uranus Obsid & Pointing offset ($^{\prime\prime}$) & Dark Sky Obsid  \\
\hline
SLWC3, SSWD4$^{\rm (a)}$ & 2010-06-01 (OD\,383) & 22 HR & 1342197472 & $0.6\pm0.3$ & 1342197456  \\
SLWC3, SSWD4$^{\rm (b)}$ & 2012-05-30 (OD\,1112) & 22 HR & 1342246285 & $1.8\pm0.3$ & 1342246266 \\
SLWC3, SSWD4$^{\rm (a)}$ & 2012-01-10 (OD\,972) & 6 LR & 1342237016 & $2.8\pm0.4$ & 1342237003   \\
SLWC3, SSWD4$^{\rm (b)}$ & 2012-05-30 (OD\,1112) & 22 LR & 1342246283 & $2.6\pm0.3$ & 1342246264 \\
\hline
SLWB2, SSWE5 & 2010-06-01 (OD\,383) & 6 HR & 1342197473 & $2.9\pm0.5$ & 1342197456 \\
SLWD2, SSWE2 & 2010-06-01 (OD\,383) & 6 HR & 1342197474 & $3.4 \pm 0.5$ & 1342197456 \\
SLWC2, SSWF3 & 2010-06-01 (OD\,383) & 6 HR & 1342197475 & $3.3\pm0.5$ & 1342197456 \\
SLWB3, SSWC5 & 2010-06-01 (OD\,383) & 6 HR & 1342197476 & $2.0\pm0.7$ & 1342197456 \\
SLWD3, SSWC2 & 2010-06-01 (OD\,383) & 6 HR & 1342197477 & $3.0\pm0.6$ & 1342197456 \\
SLWC4, SSWB3 & 2010-06-01 (OD\,383) & 6 HR & 1342197478 & $2.3\pm0.6$ & 1342197456 \\
SSWE4 & 2011-06-19 (OD\,767) & 4 HR & 1342222865 & $6.0 \pm 0.4$ & 1342222873 \\ 
SSWE3 & 2011-06-19 (OD\,767) & 4 HR & 1342222866 & $6.7 \pm 0.4$ & 1342222873 \\ 
SSWD3 & 2011-06-19 (OD\,767) & 4 HR & 1342222867 & $5.3 \pm 0.4$ & 1342222873 \\ 
SSWC3 & 2011-06-19 (OD\,767) & 4 HR & 1342222868 & $5.2 \pm 0.4$ & 1342222873 \\ 
SSWC4 & 2011-06-19 (OD\,767) & 4 HR & 1342222869 & $3.6 \pm 0.5$ & 1342222873 \\ 
SSWB2 & 2012-01-10 (OD\,972) & 4 HR & 1342237019 & $2.8 \pm 0.6$ & 1342237004 \\
SSWB4 & 2012-01-10 (OD\,972) & 4 HR & 1342237020 & $2.5 \pm 0.5$ & 1342237004 \\
SSWD2 & 2012-01-10 (OD\,972) & 4 HR & 1342237021 & $7.1 \pm 0.4$ & 1342237004 \\
SSWD6 & 2012-01-10 (OD\,972) & 4 HR & 1342237022 & $4.4 \pm 0.5$ & 1342237004 \\
SSWF2 & 2012-01-10 (OD\,972) & 4 HR & 1342237023 & $6.1 \pm 0.4$ & 1342237004 \\
\hline
\end{tabular}
\end{center}
\begin{tablenotes}[normal,flushleft]
\item $^{\rm (a)}${before OD\,1011}
\item $^{\rm (b)}${after OD\,1011}
\end{tablenotes}
\label{tab_uranus}
\end{table*}

\label{lastpage}
\end{document}